\documentclass[12pt]{article}
\usepackage{a4}
\usepackage{epsfig,psfig}
\usepackage{supercite}

\newcommand{\nc}{\newcommand}
\nc{\Gz}{\boldsymbol {G}} \nc{\Gvc}{\boldsymbol {G}^c}
\nc{\hs}{\hspace*{1mm}} \nc{\scs}{\scriptstyle}
\nc{\beq}{\begin{eqnarray}} \nc{\eeq}{\end{eqnarray}}
\nc{\rme}{}
\nc{\no}{\nonumber}
\nc{\setval}{\fmfset{wiggly_len}{1.5mm}\fmfset{arrow_len}{1.5mm}
\fmfset{arrow_ang}{13}\fmfset{dash_len}{1.5mm}\fmfpen{0.125mm}
\fmfset{dot_size}{1thick}}
\nc{\dphi}[3]{\frac{\delta #1}{\delta  \hs
\parbox{10mm}{\centerline{
\begin{fmfgraph*}(5,3)
\fmfpen{0.125mm} \fmfleft{v1} \fmfright{v2} \fmf{plain}{v2,v1}
\fmfv{decor.size=0,label=${\scs #2}$,l.dist=0.5mm}{v1}
\fmfv{decor.size=0,label=${\scs #3}$,l.dist=0.5mm}{v2}
\end{fmfgraph*}
}}}}
\nc{\ddphi}[1]{\frac{\delta^2 #1}{\delta
\parbox{10mm}{\centerline{
\begin{fmfgraph*}(5,3)
\fmfpen{0.125mm} \fmfleft{v1} \fmfright{v2} \fmf{plain}{v2,v1}
\fmfv{decor.size=0,label=${\scs 1}$,l.dist=0.5mm}{v1}
\fmfv{decor.size=0,label=${\scs 2}$,l.dist=0.5mm}{v2}
\end{fmfgraph*}
}}\,\delta
\parbox{10mm}{\centerline{
\begin{fmfgraph*}(5,3)
\fmfpen{0.125mm} \fmfleft{v1} \fmfright{v2} \fmf{plain}{v2,v1}
\fmfv{decor.size=0,label=${\scs 3}$,l.dist=0.5mm}{v1} \f
mfv{decor.size=0,label=${\scs 4}$,l.dist=0.5mm}{v2}
\end{fmfgraph*}
}}}}
\nc{\dphidouble}[3]{\frac{\delta #1}{\delta
\parbox{10mm}{\centerline{
\begin{fmfgraph*}(5,4)
\fmfpen{0.125mm} \fmfleft{v1} \fmfright{v2}
\fmf{double,width=0.2mm}{v2,v1} \fmfv{decor.size=0,label=${\scs
#2}$,l.dist=0.5mm}{v1} \fmfv{decor.size=0,label=${\scs
#3}$,l.dist=0.5mm}{v2}
\end{fmfgraph*}
}}}}
\nc{\dphiwiggly}[3]{\frac{\delta #1}{\delta
\parbox{10mm}{\centerline{
\begin{fmfgraph*}(5,4)
\fmfpen{0.125mm} \fmfleft{v1} \fmfright{v2}
\fmf{dbl_wiggly,width=0.2mm}{v2,v1} \fmfv{decor.size=0,label=${\scs
#2}$,l.dist=0.5mm}{v1} \fmfv{decor.size=0,label=${\scs
#3}$,l.dist=0.5mm}{v2}
\end{fmfgraph*}
}}}}
\def\comment#1{}
\def\beq{\begin{eqnarray}}
\def\eeq{\end{eqnarray}}

\def\dDelta{\,\dot{}\Delta\,}
\def\dDeltad{\,\dot{}\Delta\dot{}\,}
\def\Deltad{\,\Delta\dot{}\,}
\def\ddDelta{\,\ddot{}\Delta\,}
\def\Deltadd{\,\Delta\ddot{}\,}

\def\comment#1{}

\title{Perturbation Theory for Path Integrals of Stiff Polymers}
\date{}
\author{H.~Kleinert%
\footnote{ kleinert@physik.fu-berlin.de~,
http://www.physik.fu-berlin.de/\~{}kleinert}
\, and A. Chervyakov%
\footnote{ chervyak@physik.fu-berlin.de~}
~\\
Institut f\"ur Theoretische Physik, \\
 Freie Universit\"at Berlin,   \\
 Arnimallee 14, D-14195 Berlin}

\begin{document}

\maketitle

\begin{abstract}
The  wormlike chain model of stiff polymers
is
a nonlinear $\sigma$-model in one spacetime dimension
in which the ends are fluctuating freely.
This causes important differences with respect
to the presently available
theory
which exists only for periodic
and Dirichlet boundary conditions.
We modify this theory appropriately
and
show how to perform a systematic large-stiffness expansions
for all physically interesting quantities
in powers of
 $L/\xi$, where $L$ is the length and $\xi$
the
persistence length of the polymer.
This requires
special procedures for regularizing
highly divergent
Feynman integrals
which we have developed in previous work.
We show that
by
adding to the
unperturbed action a correction term
${\cal A}^{\rm corr}$,  we can calculate all Feynman
diagrams with
Green functions satisfying Neumann boundary conditions.
Our expansions yield, order by order,
properly normalized
end-to-end distribution function
 in arbitrary dimensions $d$, its
even and odd moments,
and the
two-point correlation function.

\end{abstract}
\section{Introduction }
Recently, the study of biopolymers has become a subject of
increasing interest in the research of biological materials. Forming
networks, constituent filaments play an important role in the
structure and function of living cells and other biological
entities~\cite{sackmann,howard}. Recent advances in visualizing and
manipulating single cytoskeletal filaments~\cite{actin,spectrin} and
DNA~\cite{dna,dna1} have inspired the study of conformational
characteristics of biopolymers in single-molecule
experiments~\cite{meas1,meas2}.

Biopolymers may be {\em flexible}, such as DNA, {\em semiflexible\/}
or {\em stiff\/}, such as actin, or {\em rigid}, such as
microtubuli. For a polymer with stiffness $\kappa$, the
 {\em persistence length}
of tangent-tangent fluctuations is  $\xi
= 2\kappa/(d-1 )k_{B}T$.
Typical
values of $\xi$ in biopolymers range from several nm to a few mm.
Some example are DNA, $\xi \approx 50\,{\rm nm}$~\cite{dna};
spectrin, $\xi \approx 10\,{\rm nm}$~\cite{spectrin}; actin, $\xi
\approx 17\,{\rm \mu m}$, and microtubules, $\xi \approx 5\,{\rm
mm}$~\cite{actin}.

The length $L$ of a polymer is conveniently measured in units of
$\xi$, which defines a dimensionless parameter characterizing the
inverse stiffness, or the {\em reduced length\/} $l = L/\xi$, which
we shall also call {\em flexibility\/}. For high flexibility $l$, a
polymer behaves approximately like an ideal random chain with freely
joint links~\cite{flex}. For small flexibility $l$,  is behaves like
a rigid rod. Under the influence of longitudinal compressing forces
these filaments exhibit the classical Euler
buckling~\cite{buckling}.

Most constituent filamentary biopolymers are {\em semiflexible\/}
with intermediate values of $l$. If $l$ is of the order of unity,
the stiffness is important but not overwhelming. Effects of
stiffness have been observed even for large flexibilities up to
$l\approx1000$, for instance in long strands of DNA. Stretching
these at large forces has shown significant deviations from a random
chain model~\cite{dna1}. It is therefore important to develop a
reliable theory to calculate the influence of stiffness upon
polymers for the entire range of
flexibilities~\cite{sackmann,howard}.

If a polymer has a sizable stiffness, self-avoidance effects can be
neglected due to the energetic suppression of configurations where
the filaments fold back onto themselves. An appropriate model for a
theoretical description of semiflexible polymers is the {\em
wormlike chain model} proposed by Kratky and Porod in
1949~\cite{porod}. This model is formulated most directly in terms
of path integrals. The calculation of the statistical properties has
mostly been based so far on the equivalent diffusion
equation~\cite{saito}. For a recent review see Yamakawa's
textbook~\cite{YA}, the review article by Chirikjian and
Wang~\cite{WA}, and Chapter~15 of the textbook of one of the
authors~\cite{PI}.

The wormlike chain model is the minimal model of a polymer with
arbitrary stiffness. A central feature of this model is the local
inextensibility of filaments which is mathematically accounted for
by constraining the length of all tangent vectors to unity. Being a
sequence of $d$-component unit vectors, a polymer is equivalent to a
particle moving on the surface of a unit sphere in $d$ dimensions,
which is described by a quantum-mechanical $O(d)$-symmetric
nonlinear $\sigma$-model in one spacetime dimension.

Another equivalence exists to the continuum limit of
the classical
Heisenberg chain of unit magnets,
In fact, the partition function of the lattice model
was calculated  35 years ago by Stanley~\cite{stanley} in $d=3$ dimensions.
In the magnetic context, however,
 the quantities of interest
in polymer physics were never
considered.

Because of the nonlinearity of the wormlike chain model, only a few
of the statistical properties of the stiff polymer can be calculated
analytically, the most prominent being the mean square $\langle\,
{\bf R}^{2}\,\rangle$ of the end-to-end distance ${\bf R}$, and a
number of higher moments $\langle {\bf R}^{2n}\rangle$, for $n \leq
14$. The lowest even moments have been calculated a long time ago
analytically~\cite{hermans} and numerically~\cite{numeric} by
solving recursively the diffusion equation on the unit sphere in
three dimensions. The calculations have recently been extended for
$d=3$ up to $n=13$ in Ref.~\cite{recurs} and for arbitrary $d$ up to
$n=14$ in Ref.~\cite{klei2}. By this technique it is impossible,
however, to derive the end-to-end distribution function for all
persistence lengths. To derive this central quantity of the wormlike
chain, one has to resort to perturbation schemes. So far, only
approximate but highly accurate expressions for the end-to-end
distribution of two- and three-dimensional stiff polymers were
obtained for all persistence lengths in
Refs.~\cite{recurs,klei2,klei3}.

The end-to-end distribution function $P ({\bf R};L)$ is a principal
physical quantity characterizing the statistical properties of a
single polymer of length $L$. For models like the wormlike chain
which has only short-range interactions between monomers, it gives
also the probability density of finding {\em any two monomers\/} at
spatial distance ${\bf R} = {\bf x}(s) - {\bf x} (s')$, not just the
end points, if we identify  $L$ with the distance $|s-s'|$ between
monomers measured along the chain.

In the random-chain limit of large $l$, the distribution $P ({\bf
R};L)$ is known exactly~\cite{YA,PI} and can be approximated by a
simple Gaussian. For moderately large $l$, the corrections to the
Gaussian distribution were calculated up to the order $l^{-2}$ in
three dimensions by Daniels~\cite{daniels}, and for arbitrary $d$ by
one of the authors~\cite{PI}.

For polymers close to the rod limit, all moments and the
distribution function $P({\bf R};L)$  were calculated as expansions
in powers of $l$ for $d=3$ in
Refs.~\cite{hermans,largestiff,largestiff1}. In the same limit, the
distribution function has recently been found as an infinite series
of parabolic cylinder functions for $d=2$, and of Hermite
polynomials for $d=3$ in Ref.~\cite{frey}. The generalization to $d$
dimensions is given in Chapter~15 of the textbook~\cite{PI}.

These distribution were derived from the path integral of a simple
harmonic oscillator~\cite{frey}. For large stiffness, they are in
good quantitative agreement with Monte Carlo data. They disagree,
however, with the exact expansions of Ref.~\cite{largestiff}, except
for a few lowest orders. Since the failure has a perturbative
character, it cannot be compensated by a proper normalization. The
reason is that the harmonic path integral approximates correctly the
wormlike chain model only at lowest flexibility. It certainly needs
{\em higher-order} corrections.

If we want to calculate such corrections in the path integral
formalism, the choice of physically appropriate boundary conditions
is essential. The open-end boundary conditions of the wormlike chain
model are quite tedious for the perturbative calculation of path
integrals. For a harmonic oscillator, they can be replaced by the
simpler Neumann boundary conditions as proposed in Ref.~\cite{frey}.
However, to describe the properties of a wormlike chain, the
harmonic fluctuations must be suppressed at the endpoints. This can
be done by incorporating an important correction factor into the
path integral with Neumann boundary conditions.

Calculation schemes  based on periodic boundary conditions were
considered in Refs.~\cite{pathint3,pathint4}. Path integrals for
modified wormlike chain models with softened inextensibility
condition have been suggested in
Refs.~\cite{pathint1,pathint2,pathint5,pathint6}. These models,
however, can describe the behavior of the stiff polymer only
roughly. Some conformational properties were calculated in
Refs.~\cite{other1,other2}. Additional insights into the properties
of the end-to-end distribution function of polymers with
intermediate stiffness have also been provided recently by
Refs.~\cite{minima,extension}.

Despite a number of recent developments, the theoretical
understanding of the statistical properties of a semiflexible chain
in isolation remains much less developed than for the flexible chain,
and experimental data are often interpreted within the theoretical
framework established for random chains or rigid rods, respectively.
It is therefore important to find an unified analytic approximation
procedure which will yield a reliable end-to-end distribution over the
entire range of stiffnesses. In particular, the crossover region
between the random chain properties at low stiffness and the
rigid-rod limit should be described properly.

In this paper, we develop a systematic path integral approach to
find the properties of the wormlike chain model near the rod limit.
We shall set up a perturbation theory for a large-stiffness
expansion in powers of the flexibility $l$ and show how to calculate
all expansion coefficients analytically. All moments, the radial
end-to-end distribution function, and the two-point correlation
function will be obtained in arbitrary dimensions $d$. By this
unified approach we reproduce correctly the results of
Ref.~\cite{largestiff}, and show how to go systematically
beyond the
approximate large-stiffness result of Ref.~\cite{frey}
for a polymer in $d = 3$ dimensions.

\section{Wormlike Chain Model}
In the wormlike chain model,
the polymer is described by a
smooth curve ${\bf x} (s)$ in $d$-dimensional flat space, with
the components $x^i (s)$, $i = 1,\dots,d$. The parameter $s\in
(0,L)$ is the arc length along the curve defined by $d s = \sqrt{(d
{\bf x})^2}$. In this parametrization, the
derivatives
$
 d {\bf x} (s)/d s\equiv {\bf
u}(s) $ are tangent vectors of unit length:
\begin{eqnarray}
\mid {\bf u} (s)\!\mid\,= \,1\,.\label{2.2}
\end{eqnarray}
This important property accounts for the local inextensibility of a
polymer. All vectors ${\bf u} (s)$ lie on a surface of a unit sphere
in $d$ dimensions which has an area $S_d = 2\pi^{d/2}/ \Gamma (d/2)$
and a constant scalar curvature $R = (d-1)(d-2)$.

The local curvature of the polymer is $k (s) = \mid\dot{\bf u}
(s)\!\mid$, where $\dot{\bf u} (s)\,= \,d {\bf u} (s)/d s$. The
bending energy of the wormlike chain model is quadratic in  $k(s)$:
\begin{eqnarray}
E_{\rm c}  [{\bf u}] = \frac{k_{B}T}{2\varepsilon}\int^L_0 d s\,
\dot {\bf u}^2 (s) \,, \label{2.1}
\end{eqnarray}
where we have introduced the {\em inverse stiffness} $\varepsilon =
k_{B}T/\kappa$ in units of $k_{B}T$. The flexibility $l$ is simply
related to $ \varepsilon $ by $l=\varepsilon L(d-1)/2$.

The statistical properties of the model are completely determined by
the partition function
\begin{eqnarray}
{Z} & = & S^{-1}_{d}\,\int d^d  u_b \,\delta ({\bf u}_b^2  - 1) \int
d^d u_a \,\delta ({\bf u}_a^2  - 1)\, z({\bf u}_b,{\bf u}_a)\,,
\label{2.6}\end{eqnarray} where $z({\bf u}_b,{\bf u}_a)$  is a {\em
partition function density\/} defined by the path integral
\begin{eqnarray}
z({\bf u}_b,{\bf u}_a)= e^{\varepsilon LR/8}\int^{{\bf u} (L) = {\bf
u}_b}_{{\bf u} (0) = {\bf u}_a} {\cal D}^d  u (s) \, \delta [{\bf
u}^2 (s) - 1]\exp\left(- \frac{1}{2\varepsilon}\int_{0}^{L} d s \,
\dot {\bf u}^2 (s) \right)\,.\label{2.7}\end{eqnarray} This
coincides with the path integral for the imaginary-time evolution
amplitude $\langle {\bf u}_b \,L \vert {\bf u}_a\,0 \rangle $ of a
particle on the surface of the unit sphere in $d$ dimensions which
possesses an obvious $O(d)$-symmetry. The length parameter $s$ plays
the role of the imaginary time or {\em pseudo-time}, and the bending
energy~(\ref{2.1}) in the exponent of (\ref{2.7}) is proportional to
the {\em euclidean action} ${\cal A}_{\rme}$ of a particle:
$E_{\rm c}  [{\bf u}] = k_{B}T {\cal A}_{\rme} [{\bf u}]$. We
shall often use this analogy in the sequel.

The delta-functional in the integrand of (\ref{2.7}) accounts for
the local inextensibility~(\ref{2.2}). At the endpoints $s=0,L$,
this constraint is also accounted for by the two ordinary
$\delta$-functions in Eq.~(\ref{2.6}). The restricted ordinary
integrals over the initial and final ${\bf u}$-values lead to the
{\em partition function with open ends\/}~(\ref{2.6}). The path
integral~(\ref{2.7}) for the density is calculated with  fixed ends,
i.e., with Dirichlet boundary conditions (DBC). The extra covariant
prefactor $e^{ \varepsilon LR/8}$ is necessary to reproduce the
correct
diffusion equation on the unit sphere~\cite{PI,cherv3}
whose Hamiltonian is  a  Laplace-Beltrami operator.

The principal physical quantity of a semiflexible polymer is the
distribution function
\begin{eqnarray}
P({\bf R};L) = \left\langle\,\delta^{(d)} \left({\bf R} - \int^L_0 d
s \,{\bf u} (s)\right)\, \right\rangle\,,\label{2.3}
\end{eqnarray}
with ${\bf R} = {\bf x} (L) - {\bf x} (0)$ being the end-to-end
vector. It possesses the moments
\begin{eqnarray}
\langle\, ({\bf R}^{2})^{n/2}\,\rangle =
\left\langle\,\left[\int^L_0 \int^L_0 d s \,d s'\, {\bf u} (s)\cdot
{\bf u} (s') \, \right]^{n/2} \,\right\rangle \,,
\label{2.4}\end{eqnarray}
which have been calculated exactly
 for even $n$ up to high $n$
by solving the diffusion equation
on a sphere.
From this solution we
also
know
the
rotationally invariant two-point correlation function
\begin{eqnarray}
G(s,s') = \langle\, {\bf u} (s)\cdot {\bf u} (s')\,\rangle\,
\label{2.5}
\end{eqnarray}
as being
simply
\begin{eqnarray}
G(s,s') = \exp(- |s-s'|/\xi)\,\label{2.8}
\end{eqnarray}
displaying clearly the exponential decay of tangent-tangent
correlations over the persistence length $\xi$.
We shall  All expectation
values (\ref{2.3})--(\ref{2.5}) must be calculable
from the
path
integral representation (\ref{2.6})
of the partition function
by
inserting the quantities inside the Dirac brackets into
the integrand  of
Eq.~(\ref{2.7}).

The inextensibility condition~(\ref{2.2}) makes the path integral
non-Gaussian. For an analytic treatment, we may  follow two methods.
One may either solve the inextensibility condition~(\ref{2.2})
explicitly. Then the $O(d)$-symmetry is realized nonlinearly. In
geometric language, $O(d)$ is the {\em isometry\/} of the metric on
a surface of the sphere. This method will be used in this paper to
develop the large-stiffness perturbation expansion of path
integrals, in which all physical quantities are expressed as power
series in $l$. A second method may be based on enforcing the
inextensibility condition~(\ref{2.2}) with the help of a Lagrange
multiplier. This leads to calculations of the physical quantities as
power series in $1/d$, to be presented in a forthcoming paper
\cite{cherv5}. Both
options lead to systematic approximation schemes.

\section{Large-Stiffness Expansion}
The large-stiffness expansion of the path integral
in powers of $\varepsilon\propto l$ is a special case of
weak-coupling perturbation expansions, which have recently been
developed by us for quantum-mechanical path integrals of nonlinear
$\sigma$-models with various boundary
conditions~\cite{PI,cherv1,cherv2,cherv3,cherv4}.

\subsection{Calculation with Dirichlet Boundary Conditions}

To make contact
with the methods developed
 in the previous papers, we eliminate the $ \delta $-functional
in (\ref{2.7})
by setting ${\bf u}(s)=(q^0(s), q^\mu(s),)$
with $\mu =
1,\dots,d-1$, and integrating over $ q^0(s)$, which
yields
$q^0(s)=\sigma (s) \equiv \sqrt{1- q^2 (s)}$.
The
 partition
function density becomes
\beq z(q_b, q_a)  =  \int_{\rm DBC}{\cal D}^{d-1}\,q (s)\, \sqrt{g
(q (s))}\,\exp\left\{- {1\over 2\varepsilon}\int_{0}^{L} d s
\,g_{\mu\nu} (q)\,\dot q^\mu (s) \dot q^ \nu (s) + \varepsilon
L\,{R\over 8}\right\}\,,\label{3.9}\eeq
with the metric $g_{\mu\nu} (q) = \delta_{\mu\nu} + (1 -
q^2)^{-1}\,q_\mu q_\nu$ and the determinant $g(q) = (1 - q^2)^{-1}$.

The square root in the invariant measure
can be rewritten formally as
\begin{equation}
\prod_{s}\sqrt{g (q (s))}= \exp\left\{{\frac{1}{2}\sum_s \log  g (q
(s))}\right\}= \exp\left\{{\frac{1}{2}\int_{0}^{L} d s\, \delta (s
,s) \log g (q (s))}\right\}\,,\label{3.9a}\end{equation}
where the quantity $\int^{L}_{0} d s \,\delta (s ,s ) = L \, \delta
(0)$ counts the infinite number of eigenvalues of the operator $ -
\delta_{\mu\nu} (d/ds)^2$ in the space of functions $q^\mu(s)$.
Including the exponent of (\ref{3.9a})
into the action we are left
with the partition function density
\beq
z(q_b, q_a)  =  \int_{\rm DBC}{\cal D}^{d-1}\,q (s)\,
\exp\left\{- {\cal A}_{\rme} [q]\right\},
\eeq
with the euclidean action
\begin{eqnarray}
{\cal A}_{\rme} [q] = {1\over 2\varepsilon}\int_{0}^{L} d s
\left[\,g_{\mu\nu} (q)\,\dot q^\mu (s) \dot q^ \nu (s) -
\varepsilon\delta (s ,s) \log g (q (s))\right] - \varepsilon
L\,{R\over 8}\,.\label{3.10}
\end{eqnarray}
Repeated indices are summed as usual, i.e.,
 $q_\mu q^\mu=q^2$.
In a
perturbation expansion of the path integral~(\ref{3.9}), the inverse
stiffness $\varepsilon\propto l$ plays the role of a {\em coupling
constant}. The large-stiffness or small-flexibility expansion is a
weak-coupling expansion. The subscript DBC in Eq.~(\ref{3.9})
emphasizes the Dirichlet boundary conditions $q^\mu (L) =
q^\mu_b,\,q^\mu (0) = q^\mu_a$.

To derive the perturbation expansion of
Refs.~\cite{PI,cherv1,cherv2,cherv3,cherv4}, it is convenient to
rescale the coordinates $q^\mu(s)\rightarrow\sqrt{\varepsilon}
q^\mu(s)$. This removes $\varepsilon$ from the quadratic part of the
action. The path integral for $ \varepsilon = 0$ is therefore
Gaussian and determines the basic free correlation function or {\em
propagator\/}
\begin{eqnarray}
\langle\,q^\mu (s) q^\nu (s')\,\rangle_0 = \delta^{\mu\nu}\Delta
(s,s')\,,\label{3.16}\end{eqnarray} where $\Delta (s,s')$ is the
Green function of the operator $-d^2_s$ in the space of functions $q
(s)$ with appropriate boundary conditions.

An expansion of the action in powers of $\varepsilon$ organizes the
infinitely many interaction terms. They give rise to fluctuation
corrections which are calculated by performing all possible Wick
contractions. These consist of multiple integrals over the
pseudo-time $s$ containing products of propagators (\ref{3.16}) and
their $s$-derivatives. These integrals are mathematically undefined,
since derivatives of the propagator~(\ref{3.16}) contain {\em
generalized functions\/} or {\em distributions\/}, such as Heaviside
functions $\Theta(s-s')$ and Dirac $ \delta $-functions $ \delta
(s-s')$. The problem of defining these integrals has, fortunately,
been solved in our previous work~\cite{cherv1,cherv2,cherv3}. We
have shown that all these highly singular integrals are defined {\em
uniquely} on the basis of the simple physical requirement that the
path integrals should be independent of the coordinates used to
parametrize the unit sphere. The new integration rules are valid for
path integrals with all boundary conditions. Moreover, they are in
complete agreement with much more cumbersome calculations in $D =
1-\epsilon$ dimensions, in which the limit
$\epsilon\,\rightarrow\,0$ is taken at the end ({\em dimensional
regularization}). For one-dimensional $\sigma$-models with Dirichlet
and periodic boundary conditions, our rules have led to a perfect
cancellation of all UV-divergent diagrams involving powers of $\delta
(0)$ in each perturbative order~\cite{cherv2,cherv3,cherv4}. This
permitted us to calculate finite fluctuation corrections to any
desired order in perturbation theory.

The method developed in Refs.~\cite{PI,cherv1,cherv2,cherv3,cherv4}
can be used now to derive the partition function with open
ends~(\ref{2.6}). In coordinates $q^\mu (s)$, this quantity takes
the form
\begin{eqnarray}
{Z} & = &S^{-1}_{d}\int d^{\,d-1}  q_b \,\sqrt{g (q_b)} \int
d^{\,d-1} q_a \,\sqrt{g (q_a) }\,z(q_b, q_a)\,.
\label{3.21}\end{eqnarray}
This way of writing $Z$
suggests that we must
compute the path integral~(\ref{3.9}) with
Dirichlet boundary conditions, and subsequently perform
two ordinary integrals
over initial and final $q^\mu$-values.
The advantage of
this
procedure would be that all necessary tools
are available
from our previous work
in
Ref.~\cite{PI,cherv3}. We would expand
the action (\ref{3.10}) in powers of the coordinates $q^\mu (s)$ around
the straight-line solution
connecting the end-points, and
evaluate
 higher-order fluctuation
corrections with the help of the basic propagator (\ref{3.16})
with  Dirichlet boundary conditions
\beq \Delta_{\rm D} (s,s') = - \mid s-s'\mid/2 + (s + s')/2 - s
s'/L\,.\label{@}\eeq
The resulting expansion of $z(q_b,q_a)$ involves
the geometric
invariants formed from the curvature tensor on $q$-space. For the
unit sphere in $d$ dimensions, only powers of the curvature scalar
$R =
(d-1)(d-2)$ remain, and the expansion has the form
\begin{eqnarray}\label{3.18}
z(q_b, q_a) &=& \langle q_b \,L  \vert q_a\,0 \rangle =
\frac{e^{-{\cal A}^{\rm cl}_{\rme} [{\Delta q;\,\varepsilon}]}}{
\sqrt{2\pi L }^{d}}\,\, \sum^\infty_{k=0} L^k a_k (\Delta q;
\varepsilon)\,,
\end{eqnarray}
where the classical action in the exponent
depends on
$\Delta q^\mu \equiv (q_b - q_a)^\mu$ as
\begin{eqnarray}\label{3.19}
{\cal A}^{\rm cl}_{\rme} [{\Delta q;\,\varepsilon}] &\equiv &
{(\Delta q)^2\over 2L}\, + \,\varepsilon {[(\Delta q)^2]^2\over
6L}\, + \, \dots\,,
\end{eqnarray}
and the lowest coefficients $a_k (\Delta q; \varepsilon)$ have the small-$
\varepsilon $ expansions
\begin{eqnarray}\label{3.20}
a_0(\Delta q; \varepsilon) &\equiv & 1\, + \,\varepsilon {(d-2)\over
12}(\Delta q)^2\, + \,\varepsilon^2 {(d-2)(5d-6)\over 1440} [(\Delta
 q)^2]^2 \, + \,
\dots\,,\nonumber\\
a_1 (\Delta q; \varepsilon) &\equiv & \varepsilon {(d-1)(d-2)\over
12}\, + \,\varepsilon^2 {(d-2)(5d^2 - 17d + 18)\over 720}
(\Delta q)^2 \, + \, \dots\,,\nonumber\\
a_2 (\Delta q; \varepsilon) &\equiv & \varepsilon^2 {(d-1)(d-2)(5d^2
- 17d + 18)\over 1440} \, + \, \dots\,.
\end{eqnarray}

From this we can immediately
find the partition function
(\ref{3.21}) with open
ends. Inserting (\ref{3.18}) into
(\ref{3.21}) yields
\begin{eqnarray}
{Z} & = & 1\,, \label{3.21a}\end{eqnarray}
 to all orders in $\varepsilon$.

\subsection{Calculation with Neumann Boundary Conditions}

For a calculation of the important polymer
properties (\ref{2.3})--(\ref{2.5})
this procedure, although well defined,
would be too tedious.
To save efforts, we prefer
to do the
calculations
from a path integral in which the
ordinary integrals over endpoints
in Eq.~(\ref{2.6}) are included
in the path integral
by
suitable boundary
conditions. For a simple harmonic oscillator, these
were shown in \cite{PI} to be
the {\em Neumann boundary conditions} (NBC), as anticipated
by
Ref.~\cite{frey}. These conditions may, however, not be used
directly for
calculating the fluctuation corrections since they
are physically incorrect. The Neumann
boundary conditions
 admit only fluctuations
in which the endpoints
have
zero derivatives ${\dot q}^\mu (0) ={\dot q}^\mu (L)=0$, thus neglecting
all
paths with nonzero derivatives. We shall
see that this omission can be compensated
by adding to the action an important
correction term, so that Neumann boundary conditions
may be used after derive the
 correct perturbation expansions.

We
shall evaluate represent the quantities~(\ref{2.3})--(\ref{2.5})
from
 path integrals with Neumann boundary conditions,
 by replacing the
combined measure of Eqs.~(\ref{2.6}) and (\ref{3.9}) as follows
\begin{eqnarray}
\int\!\!d^{\,d-1} q_b \sqrt{g (q_b)}\int\!\!d^{\,d-1} q_a \sqrt{g
(q_a)}\int_{{\rm DBC}}\!{\cal D}^{d-1} q (s) \rightarrow  \int_{{\rm
NBC}}\!{\cal D}^{d-1} q (s) \,J[q_b, q_a]\,,\label{r}\end{eqnarray}
where the Jacobian factor $J [q_b, q_a]$
corrects for the missing
endpoint
fluctuations when restricting the paths to zero derivatives
at the
ends. This factor may be attributed
to an extra
action at the end points
\begin{equation}
{\cal A}_{{\rme}}^{\rm cor} [q_b, q_a] = - \log J [q_b ,q_a] = -
[q^2 (0) + q^2 (L)]/4,
\label{CORRT}\end{equation}
to be  added to
(\ref{3.10}).

\comment{
Note that such a
surface term will also appear in terms of components of the tangent
vector ${\bf u} (s)$, if we express the local constraint~(\ref{2.2})
for all values of $s$, $0\leq s \leq L$, including the endpoints
$s=0$ and $s=L$, in the linear representation of the symmetry group
$O(d)$. For the partition function given in Eqs.~(\ref{2.6})
and~(\ref{2.7}) this will correspond to the replacement of the
$\delta$-functional and the two ordinary $\delta$-functions by their
Fourier representations via introducing Lagrange multipliers
$\lambda (s)$, $b$ and $a$, respectively. The partition function
represented in such an alternative way can be calculated via
$1/d$-approximation for $d$ large at $\varepsilon d$ fixed. For
details, see the forthcoming paper~\cite{cherv5}, where we found the
uniform saddle point solution $\left(\lambda^{*} (s) =
\omega^2,\,b^{*},\,a^{*}\right)$ implying that $L^{-1}\,\int^L_0 d s
\,\langle\, {\bf u}^2 (s) \,\rangle_* = 1$. To this leading
large-$d$ order the model becomes therefore harmonic with the extra
surface action which was first introduced {\em ad hoc} by authors of
Ref.~\cite{pathint2} (see also the so-called
equilibrium~\cite{pathint5} and mean-field~\cite{pathint6}
approximations).}

Our correction term
 (\ref{CORRT}) should no be
confused with a
similar-looking
but
completely unrelated
term
 $ \lambda\{ [{\bf u}^2(0)-1]+[{\bf u}^2(L)-1]\}$  in harmonic models of stiff polymers
\cite{pathint2,pathint5,pathint6}, where it
serves to enforce {\em approximately\/} the inextensibility
(\ref{2.2}) at the endpoints, which in our approach is is satisfied {\em exactly\/}
for the entire polymer.

Thus we shall calculate the polymer properties
(\ref{2.3})--(\ref{2.5})
by performing the averages
inside with respect to
the path integral representation for $Z$ with Neumann boundary
conditions
\begin{eqnarray}
Z = S^{-1}_{d}\int_{\rm NBC}\!\! \!{\cal
D}\hspace{1pt}{\hspace{1pt}}^{d-1} q (s) \exp\!\left\{-{\cal
A}_{\rme} [q] - {\cal A}_{\rme}^{\rm cor} [q_b,
q_a]\right\}\!. \label{ZFU}\end{eqnarray} The perturbation expansion
of this expression is not straightforward since with Neumann
boundary conditions, the fluctuations of the free part of the action
contain $d-1$ zero modes. These
are due to the $d-1$ constant
 solutions
$q^\mu (s)=c^\mu$ of the ``equation of motion" $-(d/ds)^2q^ \mu(s) =
0$. Their symmetry origin lies in the $d-1$ isometries of a sphere
in $d$ dimensions. They prevent us
from  inverting the
operator $-(d/ds)^2$ to obtain a perturbative
propagator~(\ref{3.16}). The zero modes must first be extracted from
the path integral.

The proper way of doing this has recently been exhibited for
one-dimensional path integrals of a nonlinear $\sigma$-model with
periodic paths in Ref.~\cite{cherv4}. The method can easily be
modified to apply to Neumann boundary conditions. To avoid
overcounting of the constant paths in the elimination process, we
have to fix a coordinate system $q^\mu (s)$, for example, by placing
the center of mass of a polymer at the origin. Let us assume that
the end-to-end distance vector ${\bf R} = \int^L_0 d s\,{\bf u} (s)$
points towards the $d$th direction. In coordinates $q^\mu (s)$, this
yields the zero average
\begin{eqnarray}
L^{-1}\int^L_0 d s\,q^\mu (s) = 0\,,\quad \mu =1,\dots,d-1\,.
\label{3.23}\end{eqnarray} We therefore introduce the measure of
fluctuations without the dangerous $d-1$ zero modes as
\begin{equation}
{\cal D}\hspace{1pt}'{\hspace{1pt}}^{d-1} q (s) = {\cal D}^{\,d-1} q
(s)\,\delta^{(d-1)}\left[\,\int^L_0 d s\,q^\mu (s)\right]
\label{@}\end{equation}
to be used instead of the measure in Eq.~(\ref{r}).

This can be done
most directly for the distribution
function~(\ref{2.3}). Introducing the {\em reduced end-to-end
distance} $r\equiv |{\bf R}|/L $, we represent this quantity via the
path integral
\begin{eqnarray}
P(r;L) = S^{-1}_{d}\int_{\rm NBC}\!\! \!{\cal
D}\hspace{1pt}'{\hspace{1pt}}^{d-1} q (s) \delta \!\left(\!r -L^{-1}
\!\!\int^L_0 d s \,\sqrt{1 - q^2 (s)}\right)\!\exp\!\left\{-{\cal
A}_{\rme} [q] - {\cal A}_{\rme}^{\rm cor} [q_b,
q_a]\right\}\,.\label{3.32}\end{eqnarray}
In turn, the partition function~(\ref{ZFU}) can be obtained from the
distribution~(\ref{3.32}) as
\begin{equation}
{Z} = ~S_{d}\int^{\infty}_0 d r\, r^{d - 1}\,P (r;L)\,,
\label{@Zpart}\end{equation}
where the prefactor $S_{d}$ reflects the rotation invariance of
$P(r;L)$. Since the zero modes are already excluded from the path
integration in Eq.~(\ref{3.32}), we integrate simply over $r$ in
Eq.~(\ref{@Zpart}) with help of the $\delta$-function in
Eq.~(\ref{3.32}) to define the path integral~(\ref{ZFU}). This
produces a further contribution to the action~(\ref{3.10}):
\begin{equation}
{\cal A}_{{\rme}}^{\rm FP}
[q] = - (d-1) \log\,\left(L^{-1}\int_0^L d s\,\sqrt{1 - q^2
(s)}\,\right),
\label{@FP0}\end{equation}
to be referred to as {\em Faddeev-Popov action\/}. For periodic
paths this action was found in our paper~\cite{cherv4}. It is a
logarithmic Jacobian which compensates the distorting effect of the
condition~(\ref{3.23}) on the measure of path integration. This
procedure of extracting the zero modes guarantees therefore the
independence of path integrals of the choice of the coordinate
system~(\ref{3.23}) used in the perturbative calculation.

Thus we rewrite
the partition function
(\ref{ZFU}) as
a path integral
\begin{eqnarray}
Z  = \int_{\rm NBC} {\cal D}\hspace{1pt}'{\hspace{1pt}}^{d-1} q
(s)\exp\left\{- {\cal A}_{\rme} [q] - {\cal A}^{\rm cor}_{{\rm
e}} [q_b, q_a] - {\cal A}^{\rm FP}_{\rme}
[q]\right\}\,,\label{3.28}\end{eqnarray}
where the paths
not only satisfy
the Neumann boundary conditions, but have also no zero modes.
Note that the prefactor $S_d$
in (\ref{ZFU}), the volume of the isometries,
 is now absent.

The expectation values in Eqs.~(\ref{2.3})--(\ref{2.5}) can now be
defined with respect to this partition function as follows
\beq \left\langle\,\dots\,\right\rangle \equiv \int_{\rm NBC} {\cal
D}\hspace{1pt}'{\hspace{1pt}}^{d-1} q
(s)\,\left(\dots\right)\,\exp\left\{- {\cal A}_{\rme} [q] - {\cal
A}^{\rm cor}_{\rme} [q_b, q_a] - {\cal A}^{\rm FP}_{\rme}
[q]\right\}\,. \label{3.28a}\eeq
Note that there is no need to normalize the path
integral~(\ref{3.28a}) by $Z=Z_0$, since the partition
function~(\ref{3.28}) is equal unity as in Eq.~(\ref{3.21a}). This
will be verified order by order in perturbation expansion of the
path integral~(\ref{3.28}).

With the definition
(\ref{3.28a}), the radial distribution function is
given by  the path integral
\beq P^{\rm rad} (r; L)\!\!\! &=& \!\!\!\left\langle\delta \!\left(\!r - L^{-1}
\!\!\int^L_0 d s \,\sqrt{1 - q^2 (s)}\right)\right\rangle\nonumber\\
& = &\!\!\! \int_{\rm NBC}\!{\cal D}\hspace{1pt}'{\hspace{1pt}}^{d-1} q
(s) \delta \!\left(\!r -L^{-1} \!\!\int^L_0 d s \,\sqrt{1 - q^2
(s)}\right)\!\exp\left\{- {\cal A}_{\rme} [q]\! -\! {\cal A}^{\rm
cor}_{\rme} [q_b, q_a]\! -\! {\cal A}^{\rm FP}_{\rme}
[q]\right\}\nonumber\\
&=&  S_{d}\,\,r^{d - 1}\,P (r;L)\,,\label{3.28b}\eeq
where the Faddeev-Popov action~(\ref{@FP0}) is illuminated by the
$\delta$-function, whereas
the moments~(\ref{2.4}) of the reduced
distance $r$ are found from
\beq \langle\,({\bf R}^{2}/L^2)^{n/2}\,\rangle\!\!\! &=&\!\!\! \langle\,
r^n\,\rangle =  \left\langle\left(L^{-1}
\!\!\int^L_0 d s \,\sqrt{1 - q^2 (s)}\right)^n \right\rangle\nonumber\\
& = & \!\!\!\int_{\rm NBC}\!{\cal D}\hspace{1pt}'{\hspace{1pt}}^{d-1} q
(s) \left(L^{-1} \!\!\int^L_0 d s \,\sqrt{1 - q^2 (s)}\right)^n
\exp\left\{- {\cal A}_{\rme} [q] \!-\! {\cal A}^{\rm cor}_{\rme}
[q_b, q_a]\! -\! {\cal A}^{\rm FP}_{\rme}
[q]\right\}\nonumber\\
&=& \!\!\!\int_{\rm NBC} {\cal D}\hspace{1pt}'{\hspace{1pt}}^{d-1} q
(s)\exp\left\{- {\cal A}_{\rme} [q] - {\cal A}^{\rm cor}_{{\rm
e}} [q_b, q_a] - {\cal A}^{\rm FP}_{\rme n} [q]\right\} \equiv
{Z}_{n}\,.\label{3.37}\eeq
This looks similar to Eq.~(\ref{3.28}) except for the last action
term which now is
\begin{equation}
{\cal A}_{{\rme} n}^{\rm FP} [q] = - (n+d-1)
\log\,\left(L^{-1}\int_0^L d s\,\sqrt{1 - q^2 (s)}\,\right)\,,
\label{@FPn}\end{equation} rather than~(\ref{@FP0}). The path
integral~(\ref{3.37}) abbreviated by $Z_n$ comprises
for $n = 0$ the unit
partition function~(\ref{3.28}): $Z_0 = Z=1$.

With the zero modes subtracted, the
 Green function with Neumann boundary conditions
representing the lines in the Feynman diagrams
acquires the form
\beq \Delta'_{\rm N} (s, s') = L/3 - \mid s-s'\mid/2 - (s + s')/2 +
(s^2 + s'^2)/2L\,,\label{3.33}\eeq
\beq
\int^L_0 d s\,\Delta'_{\rm N} (s, s') &=& 0\,.
\label{3.33b}\eeq In the following we shall simply write $\Delta (s,
s')$ for $\Delta'_{\rm N} (s, s')$, for brevity.

\section{Partition Function and All Moments up to Four Loops}
We present now the explicit perturbative calculation of the
partition function~(\ref{3.28}) and all moments~(\ref{3.37}) to
order $\varepsilon ^2\propto l^2$. This requires evaluating Feynman
diagrams up to four loops. As mentioned before, the fluctuation
corrections involve integrals over products of distributions, which
will be calculated unambiguously using the rules derived in
Refs.~\cite{cherv2,cherv3}.

For perturbation calculation, we rescale the coordinates $q^\mu (s)
\rightarrow \sqrt{\varepsilon}\,q^\mu (s)$ and rewrite the path
integral~(\ref{3.37}) as
\begin{eqnarray}
Z_{n}  = \int_{\rm NBC} {\cal D}\hspace{1pt}'{\hspace{1pt}}^{d-1} q
(s)\exp\left\{- {\cal A}^{\rm tot}_{\rme,\,n} [q; \varepsilon]
\right\}\,,\label{4.1}\end{eqnarray}
where the total action ${\cal A}^{\rm tot}_{\rme,\,n} [q;
\varepsilon] \equiv {\cal A}_{\rme} [q; \varepsilon] + {\cal
A}^{\rm cor}_{\rme} [q_b, q_a; \varepsilon] + {\cal A}^{\rm
FP}_{\rme,\,n} [q; \varepsilon]$ reads explicitly,
\beq {\cal A}^{\rm tot}_{\rme,\,n} [q; \varepsilon] &=& \int^L_0
d s\,\left[\frac{1}{2}\left({\dot q}^2 + \varepsilon\,\frac{(q \dot
q)^2}{1 - \varepsilon q^2}\right) + \frac{1}{2}\,\delta (0) \log (1
- \varepsilon
q^2)\right]\nonumber\\
& - &\sigma_{n}\log \left[\frac{1}{L}\int^L_0 d s \sqrt{1 -
\varepsilon q^2}\right] - \frac{1}{4}\,\varepsilon \left[q^2 (0) +
q^2 (L)\right] - \varepsilon L \frac{R}{8}\,.\label{4.2}\eeq
The constant $\sigma_n$ is an abbreviation for $\sigma_n \equiv n +
(d-1)$. For $n=0$ the path integral~(\ref{4.1}) must yield the
normalized partition function $Z=Z_0=1$.

At $\varepsilon = 0$, the total action~(\ref{4.2}) contains only the
kinetic term
\begin{equation}
{\cal A}^{\rm 0}_{\rme} [q]= \frac{1}{2} \int^L_0 d s\,{\dot
q}^2(s)\,.\label{@freeac}\end{equation}
With this free action, the path integral~(\ref{4.1}) becomes
Gaussian which will be defined as
\beq Z^0\equiv \int_{\rm NBC} {\cal
D}\hspace{1pt}'{\hspace{1pt}}^{d-1} q (s)\,e^{- {\cal A}^{\rm
0}_{\rme} [q]} = \int_{\rm NBC} {\cal
D}\hspace{1pt}'{\hspace{1pt}}^{d-1} q (s)\,e^{- (1/2)\int^L_0 d s
\,{\dot q}^2 (s)} = 1\label{4.6}.\eeq
To find the corrections, we split
\begin{equation}
{\cal A}^{\rm tot}_{\rme,\,n} [q; \varepsilon] = {\cal A}^{\rm
0}_{\rme} [q] + {\cal A}^{\rm int}_{\rme,\,n} [q;
\varepsilon]\,,\label{@}\end{equation}
where the interaction action ${\cal A}^{\rm int}_{\rme,\,n} [q;
\varepsilon]$ has to be expanded in powers of small $\varepsilon$.
Its large-stiffness expansion starts out as
 \beq{\cal A}^{\rm int}_{\rme,\,n} [q; \varepsilon] =
\varepsilon {\cal A}^{\rm int1}_{\rme,\,n} [q] + \varepsilon^2
{\cal A}^{\rm int2}_{\rme,\,n} [q] + \dots\,.\label{4.3}\eeq
The first expansion term is
\beq {\cal A}^{\rm int1}_{\rme,\,n} [q] &=& \frac{1}{2}\int^L_0 d
s\left\{[q (s)\dot q (s)]^2 - \rho_{n} (s)\,q^2 (s)\right\} - L
\frac{R}{8}\,,\label{4.4}\eeq
where $\rho_{n} (s) \equiv \delta_{n} + \left[\delta (s) + \delta (s
- L)\right]/2$ with $\delta_{n} = \delta (0) - \sigma_{n}/L$. The
second expansion term is given by
\beq {\cal A}^{\rm int2}_{\rme,\,n} [q] =\frac{1}{2} \int^L_0 \!d
s\left\{[q (s)\dot q (s)]^2 - \frac{1}{2}\left[\delta (0) -
\frac{\sigma_{n}}{2L}\right]q^2 (s)\right\}q^2 (s) +
\frac{\sigma_{n}}{8L^2}\int^L_0 \!d s \int^L_0 \!d s'\,q^2 (s) q^2
(s')\,.\label{4.5}\eeq

The perturbation expansion of the path integral~(\ref{4.1}) in
powers of $ \varepsilon $ is an expansion in terms of expectation
values to be calculated with respect to the Gaussian
integral~(\ref{4.6}). For an arbitrary functional $F[q]$ of $q(s)$,
these will be denoted by
\begin{equation}
\left\langle \,F[q]\,\right\rangle _0\equiv \int_{\rm NBC} {\cal
D}\hspace{1pt}'{\hspace{1pt}}^{d-1} q (s) \,F[q]\,e^{- (1/2)\int^L_0
d s \,{\dot q}^2 (s)}\,.\label{@}\end{equation}
With this notation, the perturbative expansion of~(\ref{4.1}) reads
\beq Z_{n} &=& 1 -
\langle {\cal A}_{{\rme,\, n}}^{{\rm int}} [q;\varepsilon]\rangle_0
+ \frac{1}{2} \langle {\cal A}_{{\rme,\,
n}}^{{\rm int}} [q;\varepsilon]^2\rangle_0 - \dots \nonumber\\
&=& 1 - \varepsilon\,\langle {\cal A}_{{\rme,\, n}}^{{\rm int1}}
[q]\rangle_0 + \varepsilon^2\left(- \langle{\cal A}_{{\rme,\,
n}}^{{\rm int2}}[q]\rangle_0 + \frac{1}{2}\langle{\cal A}_{{{\rm
e},\, n}}^{{\rm int1}} [q]^2\rangle_0\right) - \dots\,.\label{4.7}
\eeq
The expectation values $\langle\dots\rangle_0$ are evaluated by
performing all possible Wick contractions with the basic
propagator~(\ref{3.16}) and the Green function~(\ref{3.33}) of the
unperturbed action~(\ref{@freeac}). The relevant loop integrals
$I_i$ and $H_i$ are calculated with our regularization rules when
necessary and are listed in Appendices~\ref{APPA} and~\ref{APPB}.

We now state the results for various terms appearing in
Eq.~(\ref{4.7}): \beq \langle {\cal A}_{{\rme},\,{n}}^{{\rm
int1}} [q]\rangle_0 = \frac{(d-1)}{2}\left[\frac{\sigma_{n}}{L} I_1
+ d I_2 - \frac{1}{2}\Delta (0,0) - \frac{1}{2}\Delta (L,L)\right] -
L\frac{R}{8} = L\frac{(d-1)n}{12}\,,\label{4.9}\eeq
\beq \langle {\cal
A}_{{\rme},\, {n}}^{{\rm int2}} [q]\rangle_0 &=&
\frac{(d^2-1)}{4}\left[\left(\delta (0) + \frac{\sigma_{n}}{2L}\right) I_3 +
2(d + 2) I_4\right] + \frac{(d-1)\sigma_{n}}{8L^2}\left[(d-1)
I^2_1 + 2 I_5 \right]\nonumber\\
&=& L^3 \frac{(d^2-1)}{120}\delta (0) + L^2
\frac{(d-1)}{1440}\left[(25d^2 + 36d + 23) + n(11d + 5)\right]\,,
\label{4.10}\eeq
\beq \frac{1}{2}\langle{\cal A}_{{\rme},\,{\rm
n}}^{{\rm int1}} [q]^2\rangle_0 &=&
\frac{L^2}{2}\left\{\frac{(d-1)L}{12}\left[\left(\delta (0) +
\frac{d}{2L}\right) - \left(\delta_{n} +
\frac{2}{L}\right)\right] -
\frac{R}{8}\right\}^2\nonumber\\
&+& \frac{(d-1)}{4}\left[H^{n}_1 - 2(H^{n}_2 + H^{n}_3 -
H_5) + H_6
-4d(H^{n}_4 - H_7 - H_{10})\right.\nonumber\\
&+&\left. H_{11} + 2d^2 (H_8 + H_9)\right] +
\frac{(d-1)}{4}\left[dH_{12} + 2(d+2)H_{13} +
dH_{14}\right]\nonumber\\
&=& \frac{L^2}{2}\left[\frac{(d-1)n}{12}\right]^2
+L^3 \frac{(d-1)}{120}\delta (0)
\nonumber\\
&+&  L^2 \frac{(d-1)}{1440}\left[(25d^2 - 22d + 25) + 4(n + 4d -
2)\right]\nonumber\\
&+& L^3 \frac{(d-1)d}{120}\delta (0) + L^2
\frac{(d-1)}{720}\left(29d - 1\right)\,. \label{4.11}\eeq
Inserting these results into Eq.~(\ref{4.7}), we obtain the
partition function and all moments up to order  $\varepsilon
^2\propto l^2$:
\beq Z_{n} &=& 1 - \varepsilon
L \frac{(d-1)n}{12} + \varepsilon^2 L^2 \left[\frac{(d-1)^2 n^2}{288}
+ \frac{(d-1)(4n + 5d - 13)n}{1440}\right]
- {\cal O} (\varepsilon^3)\nonumber\\
&=& 1 - \frac{n}{6}\,l + \left[\frac{n^2}{72} + \frac{(4n + 5d -
13)n}{360(d-1)}\right]\,l^2 - {\cal O} (l^3)\,.\label{4.12}\eeq
For $n=0$ this yields the properly normalized partition function
$Z=Z_0=1$. For $n=2,4$ we obtain the known  even moments in $d$
dimensions
\begin{eqnarray}
\left\langle r^2 \right\rangle &=&1-\frac{l}{3}+\frac{1}{12}l^2+\dots~,\\
\left\langle r^4 \right\rangle &=&1-\frac{2l}{3}+
\frac{25d-17}{90(d-1)}l^2 + \dots~. \label{@}\end{eqnarray}
In addition, we find all odd moments up to order $l^2$,
the lowest being
\begin{eqnarray}
\left\langle r \right\rangle &=&1-\frac{l}{6}+\frac{5d-7}{180(d-1)}l^2+\dots~,\\
\left\langle r^3 \right\rangle &=&1-\frac{l}{2}+
\frac{25d-4}{30(d-1)}l^2+\dots~. \label{@}\end{eqnarray}

Restricted to $d = 3$ dimensions, all moments~(\ref{4.12}) are in
full agreement with the exact expansions of Ref.~\cite{largestiff}
obtained by solving the diffusion equation on a unit sphere.
\section{Correlation Function Up to Four Loops}
As an important test of this perturbation theory, we calculate the
correlation function~(\ref{2.5}) up to four loops and verify its
agreement with the exact expression~(\ref{2.8}).

Starting point is the path integral representation~(\ref{3.28a})
with Neumann boundary conditions for the two-point correlation
function
\begin{eqnarray}
G (s, s') = \langle {\bf u} (s)\cdot{\bf u} (s')\rangle = \int_{\rm
NBC} {\cal D}\hspace{1pt}'{\hspace{1pt}}^{d-1} q (s)\,U [q, q';
\varepsilon]\, \exp\left\{- {\cal A}^{\rm tot}_{\rme,\,\rm 0} [q;
\varepsilon] \right\}\,,\label{5.1}\end{eqnarray}
with the action of Eq.~(\ref{4.2}) for $n=0$. The functional in the
integrand  $U [q, q']\equiv U (q (s), q (s'))$ is the scalar product
${\bf u} (s)\cdot{\bf u} (s')$ expressed in terms of independent
coordinates $q^\mu $ as
\beq U(q(s), q(s')) \equiv {\bf u} (s)\cdot{\bf u} (s')= \sqrt{1 -
q^2 (s)}\,\sqrt{1 - q^2 (s')} + q (s)\, q (s')\,.\label{5.2}\eeq
After rescaling $q^\mu \rightarrow \sqrt{\varepsilon}\,q^\mu$ and
expanding in powers of $\varepsilon$, this becomes
\begin{equation}
U [q, q'; \varepsilon] = 1 + \varepsilon  U_1 [q, q'] +
\varepsilon^2 U_2 [q, q'] + \dots~, \label{5.2b}\end{equation}
with
\beq U_1 [q, q'] \equiv U_1 (q(s), q(s')) &=& q (s)\,q (s') -
\frac{1}{2}\,q^2 (s) -
\frac{1}{2}\,q^2 (s')\,,\label{5.3}\\
U_2 [q, q'] \equiv U_2 (q(s), q(s')) &=& \frac{1}{4}\,q^2 (s)\,q^2
(s') - \frac{1}{8}\,[q^2 (s)]^2 - \frac{1}{8}\,[q^2
(s')]^2\,.\label{5.4} \eeq
We shall attribute the integrand $U [q, q'; \varepsilon]$ to an
interaction ${\cal A}^{\rm U}_{\rme}[q; \varepsilon]$ defined by
\beq U [q, q'; \varepsilon] \equiv \exp\left(- {\cal A}^{\rm
U}_{\rme} [q; \varepsilon]\right)\,,\eeq
which has to be added to the interaction action in Eq.~(\ref{4.2})
with $n = 0$. The small $\varepsilon$-expansion, similar to
Eq.~(\ref{4.3}), reads
\beq {\cal A}^{\rm U}_{\rme} [q; \varepsilon] = - \varepsilon U_1
[q, q'] + \varepsilon^2 \left[- U_2 [q, q'] + \frac{1}{2} U_1^2 [q,
q']\right] - \dots~.\label{5.5}\eeq

Thus, the perturbation expansion of the path integral~(\ref{5.1})
takes the form
\beq G (s, s') &=& 1 - \langle ({\cal A}_{{\rme,\,\rm 0}}^{{\rm
int}} [q;\varepsilon] + {\cal A}^{\rm U}_{\rme} [q;
\varepsilon])\rangle_0 + \frac{1}{2} \langle ({\cal A}_{{{\rm
e},\,\rm 0}}^{{\rm int}} [q;\varepsilon] + {\cal A}^{\rm U}_{{\rm
e}} [q; \varepsilon])^2 \rangle_0 - \dots~.\label{5.6}\eeq
Collecting the expansions terms of Eqs.~(\ref{4.3}) and~(\ref{5.5})
in Eq.~(\ref{5.6}) and taking into account the unit normalization of
the partition function $Z = Z_0=1$ reduces this to
\beq G (s, s') = 1 + \varepsilon \langle U_1 [q, q']\rangle_0 +
\varepsilon^2 \left[\langle  U_2 [q, q']\rangle_0 - \langle U_1 [q,
q']\,{\cal A}^{\rm int1}_{\rme,\,\rm 0} [q] \rangle_0 \right] +
\dots~,\label{5.7}\eeq
where the expectation values can be calculated, as before, using the
propagator~(\ref{3.16}) with the Green function~(\ref{3.33}). For
calculation of the rotationally invariant quantity  $G (s, s') = G
(s - s')$, however, the Green function $\Delta (s, s') + C$ must be
just as good a Green function satisfying Neumann boundary conditions
as $\Delta (s, s')$.

We illustrate this explicitly by setting $C = L (a-1)/3$ with an
arbitrary constant $a$, and calculating the expectation values in
Eq.~(\ref{5.7}) using the modified Green function. Details are given
in Appendix \ref{APPC} [see Eq.~(\ref{C1})], where we list various
expressions and integrals appearing in the Wick contractions of the
expansion~(\ref{5.7}). Using these results, we find the manifestly
$a$-independent terms up to second order in $\varepsilon$:
\beq \langle U_1 [q, q']\rangle_0  =  (d - 1)\Delta_{\rm F}(s
- s') &=& - \,\frac{(d-1)}{2}\mid s - s'\mid\,,\label{5.8}\\
\langle  U_2 [q, q']\rangle_0 - \langle  U_1 [q, q']\,{\cal A}^{\rm
int1}_{\rme,\,\rm 0} [q] \rangle_0 &=& (d-1)\left[\frac{1}{2}D_1
- \frac{1}{8}(d+1)D_2^2 - K_1 -
dK_2 \right.\label{5.8b}\\
&-&\!\!\!\! \frac{(d-1)}{L}K_3+ \left.\frac{1}{2}K_4 +
\frac{1}{2}K_5 \right] = \frac{1}{8} (d - 1)^2 (s -
s')^2.\nonumber\eeq
This leads to the two-point correlation function
\beq G (s, s') &=& 1 - \varepsilon
\frac{(d-1)}{2}\mid s - s'\mid\, + \,\,\varepsilon^2
\,\frac{(d-1)^2}{8}\,(s - s')^2
- {\cal O}(\varepsilon^3)\nonumber\\
&=& 1 - \frac{\mid s - s'\mid}{\xi} + \frac{(s - s')^2}{2\xi^2} -
\dots\label{5.10}\eeq
in agreement with the small-flexibility expansion of the exact
expression~(\ref{2.8}) up to the second order in $1/\xi$.

\section{Alternative Calculation of All Moments}
It is interesting to recapitulate the result~(\ref{4.12}) by
calculating all moments~(\ref{2.4}) once more in the way used in the
previous section. To this end, we start from the path integral with
Neumann boundary conditions
\begin{eqnarray}
\langle\, r^{n}\,\rangle = \int_{\rm NBC} {\cal
D}\hspace{1pt}'{\hspace{1pt}}^{d-1} q (s)\,R^{n}[q; \varepsilon]\,
\exp\left\{- {\cal A}^{\rm tot}_{\rme,\,0} [q;
\varepsilon]\right\}\,, \label{5.11}\end{eqnarray}
where the action ${\cal A}^{\rm tot}_{\rme,\,0} [q; \varepsilon]$
is the same as in Eq.~(\ref{5.1}), while the functional $R^2 [q]$
corresponds to the square of the reduced distance $r^2$ represented
now in terms of the $d-1$ coordinates $q^\mu (s)$ with the help of
Eq.~(\ref{5.2}) as
\begin{eqnarray}
R^2 [q] = L^{-2}\int^L_0 d s\int^L_0 d s'\,\, U (q(s), q(s'))\,.
\label{5.12}\end{eqnarray}
Note that, in terms of the reduced distance $r$, the moments
(\ref{5.11}) make sense {\em initially} for all $n$.

Using Eqs.~(\ref{5.2b})--(\ref{5.4}) yields the small
$\varepsilon$-expansion
\begin{eqnarray}
R^2[q; \varepsilon] = 1 + \varepsilon R^2_1[q] + \varepsilon^2
R^2_2[q] + \dots\,, \label{5.12a}\end{eqnarray}
with
\beq R^2_1[q] &=& L^{-2}\int^L_0 d s\int^L_0 d
s'\,\, U_1 (q(s), q(s')) = - L^{-1}\int^L_0 d s\,q^2 (s)\,,\label{5.14}\\
R^2_2[q] &=& L^{-2}\int^L_0 d s\int^L_0 d s'\,\, U_2 (q(s), q(s'))\nonumber\\
& = & - (4L)^{-1} \int^L_0 d s \,[q^2 (s)]^2 + (2L)^{-2}\int^L_0 d
s\int^L_0 d s'\,q^2 (s) q^2 (s')\,,\label{5.15}\eeq
where we have used the condition~(\ref{3.23}).

The interaction associated with $R^{n} [q; \varepsilon]$ is
\beq {\cal A}^{\rm R}_{\rme} [q; \varepsilon] = - \frac{n}2 \log
R^2[q; \varepsilon] = - \varepsilon \frac{n}{2} R^2_1[q] +
\varepsilon^2 \frac{n}2 \left\{- R^2_2[q]  + \frac{1}{2}
(R^2_1[q])^2 \right\} - \dots~.\label{5.16}\eeq
Adding~(\ref{5.16}) to the action $ {\cal A}^{\rm tot}_{\rme,\,0}
[q; \varepsilon]$ leads to the perturbation expansion of the path
integral~(\ref{5.11}) up to second order in $\varepsilon$:
\beq \langle\, r^n\,\rangle &=& 1 - \langle ({\cal A}_{{{\rm
e},\,\rm 0}}^{{\rm int}} [q;\varepsilon] + {\cal A}^{\rm R}_{{\rm
e}} [q; \varepsilon])\rangle_0 + \frac{1}{2} \langle ({\cal
A}_{{\rme,\,\rm 0}}^{{\rm int}} [q;\varepsilon] + {\cal A}^{\rm
R}_{\rme} [q; \varepsilon])^2 \rangle_0 \nonumber\\
&=& 1 + \varepsilon \frac{n}{2} \langle R^2_1[q] \rangle_0 +
\varepsilon^2 \frac{n}{2} \left[\langle R^2_2[q] \rangle_0 - \langle
R^2_1[q] \,{\cal A}^{\rm int1}_{\rm e,\,\rm 0} [q] \rangle_0 +
\frac{1}{4}(n-2)\langle (R^2_1[q])^2 \rangle_0 \right]\,.
\label{5.17}\eeq
Most of the expectation values appeared here are related through
Eqs.~(\ref{5.14}) and~(\ref{5.15}) to those obtained before in
Eqs.~(\ref{5.8}) and~(\ref{5.8b}). For these, we find
\beq \langle R^2_1[q] \rangle_0\!\!& =\!\!& {1\over L^2}\int^L_0 d s
\int^L_0 d s' \langle U_1(q (s), q (s')) \rangle_0\nonumber\\
\!\!&=\!\!& - \frac{(d-1)}{2L^2}\int^L_0 d s \int^L_0 d s' \mid s -
s'\mid\,\,=\,\,- \,L \frac{(d - 1)}{6}\,,\nonumber \\\label{5.18}\\
\langle R^2_2[q] \rangle_0 - \langle R^2_1 [q]\,{\cal A}^{\rm
int1}_{\rm e,\,\rm 0} [q] \rangle_0\!\!& = \!\!& {1\over
L^2}\int^L_0 d s \int^L_0 d s' \left[ \langle U_2 (q (s), q (s'))
\rangle_0 - \langle U_1 (q (s), q (s'))\,{\cal A}^{\rm int1}_{\rm
e,\,\rm 0} [q]
\rangle_0\right]\nonumber\\
\!\!& = \!\!& \frac{(d - 1)^2}{8L^2} \int^L_0 d s \int^L_0 d s'\,(s
- s')^2 \,\,= \,\,L^2 \frac{(d - 1)^2}{48}\,.\label{5.19}\eeq
The last expectation value in Eq.~(\ref{5.17}) is calculated as
\beq
\langle (R^2_1[q])^2 \rangle_0 &=& \frac{1}{L^2}\int^L_0 d s
\int^L_0
d s'\langle q^2 (s) q^2 (s')\rangle_0 \nonumber \\
& = &  \frac{(d-1)}{L^2}\left[(d - 1)I_1^2 + 2 I_5\right] = L^2
\frac{(d - 1)(5d - 1)}{180}\,.\label{5.20}\eeq
Inserting the results~(\ref{5.18})--(\ref{5.20}) into
Eq.~(\ref{5.17}), we obtain the same perturbation expansion for all
moments up to order $l^2$ as in Eq.~(\ref{4.12}), but in the
slightly different form
\beq \langle\,r^n \,\rangle &=& 1 - \varepsilon L \frac{(d-1)n}{12}
+ \varepsilon^2 L^2 \left[\frac{(d-1)^2 n}{96} + \frac{(d-1)(5d -
1)n(n-2)}{1440}\right]
- {\cal O} (\varepsilon^3)\nonumber\\
&=& 1 - \frac{n}{6}\,l + \left[\frac{n}{24} + \frac{(5d - 1)n
(n-2)}{360(d-1)}\right]\,l^2 - {\cal O} (l^3)\,.\label{5.21}\eeq

\section{Radial Distribution up to 4 Loops}
We now turn to calculation of the most important quantity
characterizing a polymer, the distribution function~(\ref{3.32}).
Going over to the Fourier transform of the $\delta $-function, which
enforces the reduced end-to-end distance $r = L^{-1}\int^L_0 d
s\,\sqrt{1 - q^2 (s)}$ in Eq.~(\ref{3.32}), yields the Fourier
decomposition
\begin{eqnarray}
\!\!\!\!\!P (r; L) = S_d^{-1} \int{d k \over 2\pi}\, \, e^{ - i
k(r-1)}\,P (k; L)\,,\label{6.1}\end{eqnarray}
where $P (k; L)$ has to be calculated from the path integral with
Neumann boundary conditions
\begin{eqnarray}
P(k;L) = \int_{\rm NBC} {\cal D}\hspace{1pt}'{\hspace{1pt}}^{d-1} q
(s)\exp\left\{- {\cal A}^{\rm tot}_{\rme,\,k} [q; \varepsilon]
\right\}\,.\label{6.2}\end{eqnarray}
The resulting total action
\beq {\cal A}^{\rm tot}_{\rme,\,k} [q] \equiv {\cal A}_{\rme}
[q] + {\cal A}_{\rme}^{\rm cor} [q_b, q_a] - (i k/ L)\int^L_0 d s
\left(\sqrt{1 - q^2 (s)} - 1 \right)\,,\label{6.3a}\eeq
after rescaling the coordinates $q^\mu (s) \rightarrow
\sqrt{\varepsilon} q^\mu (s)$, reads explicitly
\beq {\cal A}^{\rm tot}_{\rme,\,k} [q; \varepsilon] &=&
\int^L_0 d s\,\left\{\frac{1}{2}\left[{\dot q}^2 +
\varepsilon\,\frac{(q \dot q)^2}{1 - \varepsilon q^2}\right] +
\frac{1}{2}\,\delta (0) \log (1 - \varepsilon q^2) - {ik\over L}
\left(\sqrt{1 - \varepsilon q^2} - 1\right)\right\}\nonumber\\
& - & \frac{1}{4}\,\varepsilon \left[q^2 (0) + q^2 (L)\right] -
\varepsilon L \frac{R}{8}\,\equiv\,{\cal A}^{\rm 0}_{\rme} [q] +
{\cal A}^{\rm int}_{\rme,\,k} [q; \varepsilon]\,. \label{6.3}\eeq

As before in Eq.~(\ref{4.3}), we expand the interaction in powers of
a small coupling constant~$\varepsilon$. The first term coincides
with Eq.~(\ref{4.4}), except that $\rho_n (s)$ is now replaced by
$\rho_{k} (s) = \delta_{k} + \left[\delta (s) + \delta (s -
L)\right]/2$ with $\delta_{k} = \delta (0) - ik/L$:
\beq {\cal A}^{\rm int1}_{\rme,\,k} [q] &=& \int^L_0 d
s\,\frac{1}{2}\left[(q (s)\dot q (s))^2 - \rho_{k} (s)\,q^2
(s)\right] - L \frac{R}{8}\,.\label{6.5}\eeq
The second expansion term ${\cal A}^{\rm int2}_{\rme,\,k} [q]$ is
simpler than that in Eq.~(\ref{4.5}) by not containing the last
nonlocal contribution
\beq {\cal A}^{\rm int2}_{\rme,\,k} [q] = \int^L_0 \!d
s\,\frac{1}{2}\left\{[q (s)\dot q (s)]^2 - \frac{1}{2}\left(\delta
(0) - \frac{i k}{2L}\right)q^2 (s)\right\}q^2 (s)\,.\label{6.6}\eeq
Apart from that, the perturbation expansion of the path
integral~(\ref{6.2}) has the same general form as in Eq.~(\ref{4.7})
and, therefore, reads
\beq P (k;L) & = & 1 - \langle {\cal A}_{\rme,\, {k}}^{{\rm int}}
[q;\varepsilon]\rangle_0 + \frac{1}{2} \langle {\cal A}_{{{\rm
e}},\,
{k}}^{{\rm int}} [q;\varepsilon]^2\rangle_0 - \dots \nonumber\\
&=& 1 - \varepsilon\,\langle {\cal A}_{{\rme},\,{\rm  k}}^{{\rm
int1}} [q]\rangle_0 + \varepsilon^2\left(- \langle{\cal A}_{{\rm
e},\,{k}}^{{\rm int2}}[q]\rangle_0 + \frac{1}{2}\langle{\cal
A}_{{\rme},\,{k}}^{{\rm int1}} [q]^2\rangle_0\right) -
\dots~.\label{6.7}\eeq
The expectation values can be expressed in terms of the same
integrals listed in  Appendices~\ref{APPA} and~\ref{APPB} as
follows:
\beq \langle {\cal A}_{{\rme},\,
{k}}^{{\rm int1}} [q]\rangle_0 &=&
\frac{(d-1)}{2}\left[\frac{ik}{L} I_1 + d I_2 - \frac{1}{2}\Delta
(0,0) - \frac{1}{2}\Delta (L,L)\right] - L\frac{R}{8}\nonumber\\
& = & - L\frac{(d-1)\left[(d - 1) - i
k\right]}{12}\,,\label{6.8}\eeq
\beq \langle {\cal A}_{{\rme},\, {k}}^{{\rm int2}} [q]\rangle_0
&=& \frac{(d^2-1)}{4}\left[\left(\delta (0) + \frac{i k}{2L}\right)
I_3 + 2(d + 2) I_4\right]\nonumber\\
&=& L^3 \frac{(d^2-1)}{120}\delta (0) + L^2 \frac{(d^2-1)\left[7(d +
2) + 3i k\right]}{720}\,,\label{6.9}\eeq
\beq\frac{1}{2}\langle{\cal A}_{{\rme},\,{k}}^{{\rm int1}}
[q]^2\rangle_0 &=&
\frac{L^2}{2}\left\{\frac{(d-1)L}{12}\left[\left(\delta (0) +
\frac{d}{2L}\right) - \left(\delta_{k} + \frac{2}{L}\right)\right] -
\frac{R}{8}\right\}^2\nonumber\\
&+& \frac{(d-1)}{4}\left[H^{k}_1 - 2(H^{k}_2 + H^{k}_3 -
H_5) + H_6
-4d(H^{k}_4 - H_7 - H_{10})\right.\nonumber\\
&+&\left. H_{11} + 2d^2 (H_8 + H_9)\right] +
\frac{(d-1)}{4}\left[dH_{12} + 2(d+2)H_{13} +
dH_{14}\right]\nonumber\\
&=& L^2 \frac{(d-1)^2\left[(d - 1) - i k\right]^2}{2\cdot 12^2}
\nonumber\\
&+& L^3 \frac{(d-1)}{120}\delta (0) + L^2
\frac{(d-1)}{1440}\left[(13d^2 - 6d + 21) + 4i k(2d +
i k)\right]\nonumber\\
&+& L^3 \frac{(d-1)d}{120}\delta (0) + L^2
\frac{(d-1)}{720}\left(29d - 1\right)\,.\label{6.10}\eeq
In this way, we find the large-stiffness expansion of the path
integral~(\ref{6.2}) up to order $\varepsilon^2$:
\beq \!\!\!\!\!\!\!\!\!\!\!\! P (k; L) & = & 1 + \varepsilon L
\frac{(d-1)}{12} \left[(d - 1) - i k\right] + \varepsilon^2 L^2
\frac{(d-1)}{1440}
\nonumber\\
&\times & \left[(i k)^2 (5d - 1)\right.- \left. 2i k (5d^2 - 11d +
8) + (d-1)(5d^2 - 11d + 14)\right] + {\cal O}
(\varepsilon^3)\,.\label{6.11} \eeq
The Fourier transform~(\ref{6.1}) of the expansion~(\ref{6.11})
yields the end-to-end distribution function
\beq  P (r; l) & = & S^{-1}_d \{\delta (r-1) +
\frac{l}{6}\left[\delta' (r-1) + (d - 1)\,\delta (r-1)\right] +
\frac{l^2}{360(d - 1)} \left[(5d - 1)\,\delta'' (r-1)\right.\nonumber\\
& + & \left. 2(5d^2 - 11d + 8)\,\delta' (r-1) + (d-1)(5d^2 - 11d +
14)\,\delta (r-1)\right] + {\cal O} (l^3) \}\,.\label{6.14}\eeq
This representation is convenient to calculate the moments
\beq \langle\, r^n\,\rangle = S_d \int d r\, r^{n + (d - 1)}\, P
(r;l)\,.\label{6.15}\eeq
Indeed, inserting the distribution function~(\ref{6.14}) into
Eq.~(\ref{6.15}) yields directly the moments~(\ref{4.12}) found
before by the independent calculation.

To make the contact with the results of Ref.~\cite{largestiff}
derived in $d = 3$ dimensions, we rewrite the distribution
function~(\ref{6.14}) in the form
\beq  P (r; l) = {1\over 4\pi}\left[p_0 (l)\,\delta (r-1) + p_1
(l)\, \delta' (r-1) + p_2 (l)\,\delta'' (r-1) + p_3 (l)\,\delta'''
(r-1) + \dots \right]\,,\label{6.16}\eeq
where the coefficients $p_i (l)$ are expanded for $d = 3$ up to
order $l^3$
\beq p_0 (l)& = & 1 + {1\over 3}\,l + {13\over 180}\,l^2 + {38\over
2^3\cdot 315}\,l^3 + \dots = 1 + {2\over 3}\,t + {13\over 45}\,t^2 +
{38\over 315}\,t^3 + \dots~,
\label{6.16a}\\
p_1 (l)& = & {1\over 6}\,l + {1\over 18}\,l^2 + {34\over 2^3\cdot
315}\,l^3 + \dots = {1\over 3}\,t + {2\over 9}\,t^2 + {34\over
315}\,t^3 + \dots~,
\label{6.16b}\\
p_2 (l)& = & {7\over 360}\,l^2 + {1\over 2^3\cdot 21}\,l^3 + \dots =
{7\over 90}\,t^2 + {1\over 21}\,t^3 + \dots~
\label{6.16c}\\
p_3 (l)& = & {31\over 2^3\cdot 1890}\,l^3 + \dots = {31\over
1890}\,t^3 + \dots~.\label{6.16d}\eeq
Inserting $l = 2t$ in the last lines of
Eqs.~(\ref{6.16a})--(\ref{6.16d}) shows the agreement with the
end-to-end distribution function $G (r; t)$ of
Ref.~\cite{largestiff} reduced to the form~(\ref{6.16}) after
re-expansion in Appendix~\ref{APPD}.

Let us make now the contact with the distribution functions of
Ref.~\cite{frey}. To this end, the expansion~(\ref{6.11}) can be
reassembled as
\beq P (k; L) = P_{0} (k;L)\left\{1 + \frac{(d - 1)}{6}\,l +
\left[\frac{(d - 3)}{180(d - 1)}\,i k
 + \frac{(5d^2 - 11d + 14)}{360}\right]\,l^2
+ {\cal O} (l^3)\right\}\,,\label{6.12}\eeq
where we have factored out the series containing the expansion terms
with the same powers of $\varepsilon$, $k$ and $L$,
\begin{eqnarray}
P_{0}(k;L) = 1 - \frac{(d-1)}{ 2^2 \cdot 3} (i k \varepsilon L)  +
\frac{(d-1)(5d - 1)}{2^5 \cdot 3^2 \cdot 5} (i k \varepsilon L)^2 -
\dots~.\label{6.12a}
\end{eqnarray}
By setting ${\hat\omega}^2 \equiv i k \varepsilon L$, we identify
this with the beginning of the expansion of a functional determinant
\beq P_{0}(\hat\omega) =
\left(\frac{\hat\omega}{\sinh\hat\omega}\right)^{(d-1)/2} =\,\,\,\,1
- \frac{d-1}{12}\hat  \omega ^2 +\frac{(d-1)(5d-1)}{1440} \hat
\omega ^4 + \dots~,\label{6.12b}\eeq
valid for $\hat\omega$ small. One can replace, however, the
expansion~(\ref{6.12a}) by the exact expression~(\ref{6.12b}) for
$P_0 (\hat\omega)$ and integrate over $\hat\omega$ in the Fourier
transform~(\ref{6.1}). This would ignore the  smallness of
$\hat\omega$. In such a way, the distribution function with the only
determinant~(\ref{6.12b}) was calculated for  $d =2,3$ in
Ref.~\cite{frey} from the $\hat\omega$-integral
\begin{eqnarray}
P_0 (r;l)\propto {1\over l}\,\int_{-i\infty}^{i\infty}\,{d \hat
\omega \over 2\pi i}\,e^{- \hat \omega ^2 (d-1)(r-1)/2l}\,\hat\omega
P_{0}(\hat \omega)\,,\label{6.13a}\end{eqnarray}
which was found to give, after a proper normalization, good radial
end-to-end distribution functions for large stiffness. It represents
the leading contribution to the distribution function~(\ref{6.1}),
since $P_0 (\hat\omega)$ enters Eq.~(\ref{6.12}) as a prefactor with
no dependence on the flexibility $l$. To find corrections, the rest
of expansion~(\ref{6.12}) in powers of small $l$ must be known for
all $\hat\omega$. This can be done by reorganizing the perturbation
theory as follows.

The determinant~(\ref{6.12b}) can directly be obtained from the path
integral of a simple harmonic oscillator with Neumann boundary
conditions~\cite{PI}:
\beq Z^0_\omega \equiv \int_{\rm NBC} {\cal
D}\hspace{1pt}'{\hspace{1pt}}^{d-1} q (s)\,e^{- {\cal A}^{\rm
0}_{\rme,\,\omega} [q]} = P_{0} (\hat\omega)\,,\label{6.13}\eeq
where ${\cal A}^{\rm 0}_{\rme,\,\omega} [q]$ is the harmonic
action
\begin{equation}
{\cal A}^{\rm 0}_{\rme,\,\omega} [q] = \frac{1}{2} \int^L_0 d
s\,\left[{\dot q}^2 (s) + \omega^2 q^2
(s)\right]\,.\label{harmac}\end{equation}
With the redefinition $\omega^2 \equiv i k \varepsilon /L$, it
represents a ''free'' part of the total action~(\ref{6.3}), whose
expansion in powers of small $\varepsilon$ becomes
\begin{equation}
{\cal A}^{\rm tot}_{\rme,\,k} [q; \varepsilon] = {\cal A}^{\rm
0}_{\rme,\,\omega} [q] + \varepsilon {\cal A}^{\rm int1}_{{\rm
e},\,\omega} [q] + \dots~\,,\label{6.17}\end{equation}
while the first expansion term reads
\beq {\cal A}^{\rm int1}_{\rme,\,\omega} [q] &=& \int^L_0 d
s\left [{1\over 2}\,(q \dot q )^2 - {1\over 2}\,\delta (0)\,q^2 +
{1\over 8}\,\omega^2 (q^2)^2\right] - {1\over 4}\,\left[q^2 (0) +
q^2 (L)\right] - L \frac{R}{8}\,.\label{6.18}\eeq
Note that the interaction~(\ref{6.18}) contains the harmonic terms
even at lowest order in $\varepsilon$.

The result~(\ref{6.13}) represents the leading term in the
large-stiffness expansion for the path integral~(\ref{6.2}). By
expanding the path integral~(\ref{6.2}) with respect to the harmonic
integral~(\ref{6.13}), we find the higher-order fluctuation
corrections. This yields
\beq P (\hat\omega; L) &=& Z^0_\omega\,\left(1 -
\varepsilon\,\langle {\cal A}_{{\rme,\,\omega}}^{{\rm int1}}
[q]\rangle_0 + \dots \right)\,,\label{6.19}\eeq
where the expectation values are evaluated using the basic
propagator~(\ref{3.16}), which contains now a harmonic Green
function of the unperturbed action~(\ref{harmac}). With the
zero-mode subtracted, this Green function explicitly reads
\begin{eqnarray}
\Delta_{\omega} (s, s') & = & {2 \over
L}\sum^{\infty}_{n=1}\frac{\cos (n\pi s/L)\cos (n\pi s'/L)}
{\omega^2 + n^2 \pi^2/L^2}\nonumber\\
& = & \frac{\cosh\omega(L - \mid s - s'\mid) + \cosh\omega (L - (s +
s'))}{2\omega\sinh\omega L} - {1\over L\omega^2}\,.
\label{6.20}\end{eqnarray}

As a result, we obtain the next-to-leading contribution
\beq \langle {\cal A}_{{\rme},\,\omega}^{{\rm int1}} [q]\rangle_0
& = & - L\,\frac{(d-1)^2}{8} -
L\,\frac{(d-1)}{4}\left(\frac{\coth\hat\omega}{\hat\omega} -
\frac{1}{\hat\omega^2}\right)\nonumber\\
& + & L\,\frac{(d^2-1)}{32}\left(\coth^2 \hat\omega +
\frac{\coth\hat\omega}{2 \hat\omega} - \frac{3}{2 \sinh^2
\hat\omega}\right)\,.\label{6.21}\eeq
Substituting now the leading~(\ref{6.13}) and the
next-to-leading~(\ref{6.21}) terms into expansion~(\ref{6.19})
yields
\beq P (\hat\omega; l) = P_{0}(\hat\omega)\,R (\hat\omega;
l)\,,\label{6.22}\eeq
where the factor $R (\hat\omega; l)$ for all $\hat\omega$ has the
small-$l$ expansion
\begin{equation}
R (\hat\omega; l) = 1 + l\,\frac{(3d - 5)}{16} - l\,\frac{(d +
1)}{32}\,\left(\frac{\coth\hat\omega}{\hat\omega} - \frac{1}{\sinh^2
\hat\omega}\right) + l\,{1\over
2}\left(\frac{\coth\hat\omega}{\hat\omega} -
\frac{1}{\hat\omega^2}\right) + {\cal O} (l^2)\,.
\label{6.23}\end{equation}

Expanding~(\ref{6.22}) in powers of $\hat\omega$ for $\hat\omega$
small yields, of course, the expansion~(\ref{6.12}) to the order of
$l$. Nevertheless the analytic form~(\ref{6.22}) is convenient for
the use in the Fourier representation~(\ref{6.1}) in terms of
$\hat\omega$:
\begin{eqnarray}
P(r;l) = {(d-1)\over l}\,S^{-1}_d\,\int_{-i\infty}^{i\infty}\,{d
\hat \omega \over 2\pi i}\, e^{- \hat\omega ^2 (d-1)(r -
1)/2l}\,\hat\omega P (\hat\omega; l)\,.\label{6.24}\end{eqnarray}

An important observation for performing the
$\hat\omega$-integral~(\ref{6.24}) is now that the
integrand~(\ref{6.22}) can be expressed in terms of $P_0
(\hat\omega)$, and the derivatives $P_0 ' (\hat\omega)$ and $P_0 ''
(\hat\omega)$ with respect to $\hat\omega$. To show this, we set
\beq P_{0}(\hat\omega) = \left( \frac{ \hat\omega }{\sinh \hat\omega
}\right)^{(d-1)/2} = \,\,\, e^{- f (\hat\omega)}\,,\label{6.25}\eeq
with
\beq f(\hat\omega) = {(d - 1)\over 2}\,\left(\log\sinh\hat\omega -
\log\hat\omega\right)\,.\label{6.26}\eeq
In terms of $f (\hat\omega)$, the factor~(\ref{6.23}) becomes
\begin{equation}
R (\hat\omega; l) = 1 + \frac{(3d - 5)l}{16} - \frac{(d - 15)l}{16(d
- 1)}\,\frac{f' (\hat\omega)}{\hat\omega}  - {(d + 1)l\over 16 (d -
1)}\, f'' (\hat\omega)\,. \label{6.27}\end{equation}
Substituting Eqs.~(\ref{6.25}) and~(\ref{6.27}) into
Eq.~(\ref{6.22}), we use the identities
\beq - f'(\hat\omega)\,e^{- f(\hat\omega)} & = & P_0 '
(\hat\omega)\,,\nonumber\\
- f''(\hat\omega)\,e^{- f(\hat\omega)} & = & {2\over (d + 1)}\,P_0
'' (\hat\omega) - {2(d - 1)\over (d + 1) \hat\omega}\,P_0 '
(\hat\omega) - {(d - 1)^2 \over 2(d + 1)}\,P_0
(\hat\omega)\,.\label{6.28}\eeq
This brings, finally, the expansion~(\ref{6.22}) in the form
\begin{equation}
P (\hat\omega; l) = \left[1 + \frac{(5d - 9)l}{32}\right]\,P_0
(\hat\omega) - \frac{(d + 13)l}{16(d - 1)\hat\omega}\,P_0 '
(\hat\omega) + {l\over 8(d - 1)}\, P_0 '' (\hat\omega)\,.
\label{6.29}\end{equation}

With the representation~(\ref{6.29}), the integration of
Eq.~(\ref{6.24}) is straightforward. We substitute the binomial
series
\begin{eqnarray}  \!\!\!\!\!\!\!\!\!
P_0 (\hat\omega) = \left( \frac{ \hat\omega }{\sinh \hat\omega
}\right)^{(d-1)/2} = \,\,\,(2\hat\omega)^{(d-1)/2}\sum _{k=0}^\infty
(-1)^k \left(-(d-1)/2\atop k\right)e^{-(2k + (d-1)/2)
\hat\omega}\,\label{6.30}\end{eqnarray}
and express the resulting integrals in terms of a parabolic cylinder
functions. In the physical case of three dimensions, where
$\hat\omega^2 = i k l$, the answer reads
\beq P (r; l) & = & \frac{1}{4\pi \sqrt{\pi}\,l}\sum^{\infty}_{n =
1}\,\frac{1}{\left[(1 - r)/l\right]^{3/2}}\,\exp\left[- \frac{(n -
1/2)^2}{(1-r)/l}\right]\nonumber\\
&\times&\left\{\left[1 + {(n^2 - n + 1)l\over 4}\right]\,H_2\!
\left[\frac{n - 1/2}{\sqrt{(1 - r)/l}}\right] + \frac{3(n - 1/2)^2\,
l}{\sqrt{2}} - (1 - r)\right\}\,,\label{6.31}\eeq
where the second Hermite polynomial is $H_2 (x) = 4 x^2 - 2$. The
first term in the brackets represents the leading contribution found
before in Ref.~\cite{frey}, the other are the new \mbox
{next-to-leading} corrections. The proper normalization is ensured
automatically in each perturbative order $l$ in the
small-flexibility expansion. Since the next-to-leading terms are
suppressed as $l \sim 1/\kappa$, there is no drastic change in the
behavior of the radial distribution with respect to the leading
contribution.

\section{Conclusion}
In conclusion, we have developed a systematic
 perturbation theory capable
of yielding the polymer properties near the rod limit from a path
integral formulation. This has yielded the large-stiffness
expansions for the experimentally relevant end-do-end distribution,
all the even and odd moments, and the correlation function in $d$
dimensions as power series in the flexibility $l$ up to the order
$l^2$. All these results have been obtained without the use of
the diffusion equation on a unit sphere.
In subsequent work we shall use
variational perturbation theory \cite{PI,KS}
to extend the results from small to large flexibility, i.e.,
to find results connecting the above results
smoothly with the random-chain limit.

~\\
~\\
{\LARGE{\bf Appendices}}
\appendix

\section{Basic Integrals}
\label{APPA}
The following well-defined integrals appear throughout the calculation
of the Feynman diagrams:
\beq \Delta (0,0) &=& \Delta (L,L) = \frac{L}{3}\,,
\label{A1}\\
I_{1} &=& \int_0^L d s \,\Delta (s, s) = \frac{L^2}{6}\,,
\label{A2}\\
I_{2} &=& \int_0^L  d s\,\dDelta^2 (s, s) = \frac{L}{12}\,,
\label{A3}\\
I_{3} &=& \int_0^L  d s\,\Delta^2 (s, s) = \frac{L^3}{30}\,,
\label{A4}\\
I_{4} &=& \int_0^L  d s\,\Delta (s, s)\,\dDelta^2 (s, s) =
\frac{7L^2}{360}\,,
\label{A5}\\
I_{5} &=& \int_0^L  d s  \int_0^L  d s'\,\Delta^2 (s, s') =
\frac{L^4}{90}\,,
\label{A6}\\
\Delta (0,L) &=& \Delta (L,0) = - \frac{L}{6}\,,
\label{A7}\\
I_{6} &=& \int_0^L  d s\,\left[\Delta^2 (s, 0) + \Delta^2 (s,
L)\right] = \frac{2L^3}{45}\,,
\label{A8}\\
I_{7} &=& \int_0^L  d s  \int_0^L  d s'\,\Delta (s, s)\dDelta^2 (s,
s') = \frac{L^3}{45}\,,
\label{A9}\\
I_{8} &=& \int_0^L  d s  \int_0^L  d s'\,\dDelta (s, s)\Delta (s,
s')\dDelta (s, s') = \frac{L^3}{180}\,,
\label{A10}\\
I_{9} &=& \int_0^L  d s \,\Delta (s, s)\left[\dDelta^2 (s, 0) +
\dDelta^2 (s, L)\right] = \frac{11L^2}{90}\,,
\label{A11}\\
I_{10} &=& \int_0^L  d s \,\dDelta (s, s)\left[\Delta (s, 0)\dDelta
(s, 0) + \Delta (s, L)\dDelta (s, L)\right] = \frac{17L^2}{360}\,,
\label{A12}\\
\dDelta (0,0) &=& - \dDelta (L,L) = - \frac{1}{2}\,,
\label{A13}\\
I_{11} &=& \int_0^L  d s  \int_0^L  d s'\,\Delta (s, s)\Deltad (s,
s')\dDelta (s', s') = \frac{L^3}{360}\,,
\label{A14}\\
I_{12} &=& \int_0^L  d s  \int_0^L  d s'\,\Delta (s, s')\dDelta^2
(s, s') = \frac{L^3}{90}\,. \label{A15}\eeq

\section{Loop Integrals}
\label{APPB}

We list here the Feynman integrals evaluated with dimensional
regularization rules whenever necessary. In the calculation, they
occur either from the expectations~(\ref{4.9})--(\ref{4.11}), or
from the expectations~(\ref{6.8})--(\ref{6.10}). Accordingly, we
encounter the integrals depending either on $\rho_{n} (s) =
\delta_{n} + \left[\delta (s) + \delta (s - L)\right]/2$ with
$\delta_{n} = \delta (0) - \sigma_{n}/L$, or on $ \rho_{k} (s) =
\delta_{k} + \left[\delta (s) + \delta (s - L)\right]/2$ with
$\delta_{k} = \delta (0) - ik/L$:
\beq \,\hspace{-2.2cm}\hspace{0mm}\raisebox{-1.2mm}{\mbox{\input 0dotdot.tps }}\!~~\,H^{{n}({k})}_1 & = &
\int_0^L \!d s \int_0^L \!d s' \rho_{{n}({k})} (s)\rho_{{n}({k})}
(s')\Delta^2 (s, s')\nonumber\\
& = & \delta^2_{{n}({k})} \,I_5 + \delta_{{n}({\rm
k})}\,I_6 + \frac{1}{2}\left[\Delta^2
(0,0) + \Delta^2 (0,L)\right],\label{B1}\\
~~\hspace{-2.2cm}\hspace{0mm}\raisebox{-1.2mm}{\mbox{\input 6dot.tps }}\,H^{{n}({k})}_{2} & = &
\int_0^L d s \int_0^L  d s' \rho_{{n}({k})} (s')\Delta (s, s)
\dDelta^2 (s, s') =
\delta_{{n}({k})}\,I_7 + \frac{I_9}{2},\label{B2}\\
\,\hspace{-2.2cm}\hspace{0mm}\raisebox{-1.2mm}{\mbox{\input 6pdot.tps }}\!~~\,H^{{n}({k})}_{3} & = &
\int_0^L d s \int_0^L  d s' \rho_{{n}({k})} (s') \dDeltad (s, s)
\Delta^2 (s, s') =
\left[\delta_{{n}({k})}\,I_5 + \frac{I_6}{2}\right]\delta (0)\label{B3},\\
~~\hspace{-2.2cm}\hspace{0mm}\raisebox{-1.6mm}{\mbox{\input infdoth.tps }}\,H^{{n}({k})}_{4} & = &
\int_0^L d s \int_0^L  d s' \rho_{{n}({k})} (s') \dDelta (s, s)
\dDelta (s, s') \Delta (s, s') = \delta_{{n}({k})}\,I_8 +
\frac{I_{10}}{2}\,.\label{B4} \eeq
When calculating $Z_{n}$, we need to insert here $\delta_{n} =
\delta (0) - \sigma_{n}/L$, thus obtaining
\beq H^{n}_1 & = &
\frac{L^4}{90}\,\delta^2 (0) + \frac{L^3}{45}\,(3 - d - n)\,\delta
(0) + \frac{L^2}{360}\,\left[(45
- 24d + 4d^2) - 4n(6 -2d - n)\right]\label{B1a},\\
H^{n}_{2} & = &
\frac{L^3}{45}\,\delta (0) + \frac{L^2}{180}\,(15 - 4d - 4n)\label{B2a},\\
H^{n}_{3} & = & \frac{L^4}{90}\,\delta^2 (0) +
\frac{L^3}{90}\,(3 - d - n) \delta (0)\label{B3a},\\
H^{n}_{4} & = & \frac{L^3}{180}\,\delta (0) +
\frac{L^2}{720}\,(21 - 4d - 4n)\,,\label{B4a} \eeq
where the values for $n = 0$ correspond to the partition function $Z
= Z_{\rm 0}$. The substitution $\delta_{k} = \delta (0) - ik/L$,
required for the calculation of $P (k; L)$, yields
\beq H^{k}_1 & =
& \frac{L^4}{90}\,\delta^2 (0) + \frac{L^3}{45}\,\left[2 + (- i
k)\right]\,\delta (0) + \frac{L^2}{90}\,( - i k)\left[4
+ (- i k)\right] + \frac{5L^2}{72},\label{B1b}\\
H^{k}_{2} & = & \frac{L^3}{45}\,\delta (0) +
\frac{L^2}{180}\,\left[11 + 4 (- i k)\right],\label{B2b}\\
H^{k}_{3} & = & \frac{L^4}{90}\,\delta^2 (0) +
\frac{L^3}{90}\,\left[2 + ( - i k)\right]\,\delta (0),\label{B3b}\\
H^{k}_{4} & = & \frac{L^3}{180}\,\delta (0) +
\frac{L^2}{720}\,\left[17 + 4( - i k)\right].\label{B4b}\eeq

The other loop integrals are
\footnotesize
\beq
\hspace{1pt}\hspace{-3cm}\hspace{0mm}\raisebox{-1.2mm}{\mbox{\input 8.tps }}\hspace{-1pt}~~~&&\!\!\!\!\!\!\!\!\!\!\!\!\!\!\!\!\!H_5
= \int_0^L  d s \int_0^L d s' \Delta (s, s) \dDelta^2 (s, s')
\dDeltad (s', s')= \delta (0) I_7
= \frac{L^3}{45}\,\delta (0) \label{B5},\\
\hspace{-3cm}\hspace{0mm}\raisebox{-1.2mm}{\mbox{\input 10.tps }}~~~&&\!\!\!\!\!\!\!\!\!\!\!\!\!\!\!\!\!H_{6}
= \int_0^L  d s \int_0^L d s' \dDeltad (s, s) \Delta^2 (s, s')
\dDeltad (s', s') = \delta^2 (0)
I_5  = \frac{L^4}{90}\,\delta^2 (0),\label{B6}\\
\hspace{-1pt}\hspace{-3cm}\hspace{0mm}\raisebox{-3mm}{\mbox{\input thr1hh.tps }}\hspace{1pt}~~~&&\!\!\!\!\!\!\!\!\!\!\!\!\!\!\!\!\!H_{7}
= \int_0^L d s \int_0^L d s' \dDelta (s, s) \dDelta (s, s') \Delta
(s, s') \dDeltad (s', s') =
\delta (0) I_8  = \frac{L^3}{180} \delta (0) \label{B7},\\
\hspace{-3cm}\hspace{0mm}\raisebox{-3mm}{\mbox{\input thr1hhhh.tps }}~~~&&\!\!\!\!\!\!\!\!\!\!\!\!\!\!\!\!\!H_{8}
= \int_0^L  d s \int_0^L d s' \dDelta (s, s) \dDelta (s, s') \Deltad
(s, s') \dDelta (s', s')
= - \frac{L^2}{720},\label{B8}\\
\hspace{-3cm}\hspace{0mm}\raisebox{-3mm}{\mbox{\input thr1hhh.tps }}~~~&&\!\!\!\!\!\!\!\!\!\!\!\!\!\!\!\!\!H_{9}
= \int_0^L  d s \int_0^L d s' \dDelta (s, s) \Delta (s, s') \dDeltad
(s, s') \dDelta (s', s') = \frac{I_{10}}{2} - \frac{I_8}{L} - H_8 =
\frac{7L^2}{360}\label{B9},\\
\hspace{-3cm}\hspace{0mm}\raisebox{-2.5mm}{\mbox{\input thr1h.tps }}~~~&&\!\!\!\!\!\!\!\!\!\!\!\!\!\!\!\!\!H_{10}
= \int_0^L  d s \int_0^L d s' \Delta (s, s) \dDelta (s, s') \dDeltad
(s, s') \dDelta (s', s') = \frac{I_{9}}{4} - \frac{I_7}{2L} =
\frac{7L^2}{360},\label{B10}\\
\hspace{2pt}\hspace{-3cm}\hspace{0mm}\raisebox{-1mm}{\mbox{\input 9.tps }}\hspace{-2pt}~~~&&\!\!\!\!\!\!\!\!\!\!\!\!\!\!\!\!\!H_{11}
= \int_0^L  d s \int_0^L d s' \Delta (s, s) \dDeltad^2 (s, s')
\Delta (s', s') = \delta (0) I_3 + (\frac{I_{1}}{L})^2 -
\frac{2(I_3 - I_{11})}{L} - 2I_4\nonumber\\
&&\!\!\!\!\!\! +\, 2\left[\Delta^2 (L,L)\Deltad (L,L) - \Delta^2
(0,0) \Deltad (0,0)\right] - 2H_{10} = \frac{L^3}{30} \delta (0) +
\frac{L^2}{9},\label{B11}\\
\hspace{-3cm}\hspace{0mm}\raisebox{-2mm}{\mbox{\input 13.tps }}~~~&&\!\!\!\!\!\!\!\!\!\!\!\!\!\!\!\!\!H_{12}
= \int_0^L  d s \int_0^L
d s' \dDelta^2 (s, s') \Deltad^2 (s, s')  = \frac{L^2}{90},\label{B12}\\
\hspace{-3cm}\hspace{0mm}\raisebox{-1.8mm}{\mbox{\input 12.tps }}~~~&&\!\!\!\!\!\!\!\!\!\!\!\!\!\!\!\!\!H_{13}
= \int_0^L  d s \int_0^L d s' \Delta (s, s') \dDelta (s, s') \Deltad
(s, s') \dDeltad (s', s) = \frac{I_{4}}{2} - \frac{I_{12}}{2L} -
\frac{H_{12}}{2} =
- \frac{L^2}{720},\label{B13}\\
\hspace{-3cm}\hspace{0mm}\raisebox{-2mm}{\mbox{\input 11.tps }}~~~&&\!\!\!\!\!\!\!\!\!\!\!\!\!\!\!\!\!H_{14}
= \int_0^L  d s \int_0^L d s' \Delta^2 (s, s') \dDeltad^2 (s, s') =
\delta (0) I_3 - \frac{2(I_3 - I_{12})}{L} - 2I_4 +
\frac{I_{5}}{L^2}\nonumber,\\
&&\!\!\!\!\!\! +\, 2\left[\Delta^2 (L,L)\Deltad (L,L) - \Delta^2
(0,0) \Deltad (0,0)\right] - 2H_{13} = \frac{L^3}{30} \delta (0) +
\frac{11L^2}{72}.\label{B14}\eeq
\normalsize
Calculating the integrals~(\ref{B9})--(\ref{B14}) required the
regularization rules of Refs.~\cite{cherv2,cherv3}. To compute these
unambiguously, we must first use partial integrations together
with Neumann boundary conditions so that we can
apply
subsequently the equation $\ddDelta (s, s') = \Deltadd (s, s') = 1/L
- \delta (s, s')$. In this way, most of the integrals
can  be expressed
in terms of the basic integrals
in \ref{APPA}. For the integrals~(\ref{B11})
and~(\ref{B14}), the above procedure had to be applied twice.

\section{Integrals Involving Green Functions with Arbitrary Constant}
\label{APPC} To demonstrate the translational invariance of results
showed in the main text we use the modified Green function
\beq \Delta (s,s') = \frac{L}{3}\,a - \frac{\mid s-s'\mid}{2} -
\frac{(s + s')}{2} + \frac{(s^2 + s'^2)}{2L}\,,\label{C1}\eeq
with an arbitrary constant $a$. The various relations fulfilled by
this Green function are listed below assuming $s \geq s'$, for
brevity.

The useful relations are
\beq\!\!\!\!\!\!\!\!\! \Delta_{\rm F} (s, s')\!\! &=&\!\!
\Delta_{\rm F} (s - s') = \Delta (s, s') - \frac{1}{2}\Delta (s, s)
- \frac{1}{2}\Delta (s', s') = -
\frac{1}{2}\,(s - s'),\label{C2}\\
\!\!\!\!\!\!\!\!\!D_1 (s, s')\!\! &=& \!\!\Delta^2 (s, s') - \Delta (s, s)\Delta (s', s') = (s
- s')\!\!\left[\frac{s(L - s)}{L} + \frac{(s - s')(s + s')^2}{4L^2}
- \frac{La}{3}\right]\label{C3},\\
\!\!\!\!\!\!\!\!\!D_2 (s, s')\!\! &=& \!\!\Delta (s, s) - \Delta (s', s') = \frac{(s -s')(s +
s' - L)}{L}.\label{C4}\eeq

The useful integrals are
\beq \hspace{-7mm} J_1 (s, s')\!\!\!\! &=&\!\!\!\! \int^L_0 d
t\,\Delta (t, t)\dDelta (t, s)\dDelta (t, s') = \frac{(s^4 +
s'^4)}{4L^2} -
\frac{(2s^3 + s'^3)}{3L}\nonumber\\
\hspace{-7mm}
&+&\!\!\!\! \frac{((a+3)s^2 + as'^2)}{6}- \frac{s a}{3} L + \frac{(20a-9)}{180}L^2,\label{C5}\\
J_2 (s, s')\!\!\!\! &=& \!\!\!\!\int^L_0 d t\,\dDelta (t, t)\Delta (t, s)\dDelta (t,
s') = \frac{(- 2s^4 + 6s^2 s'^2 +3s'^4)}{24L^2}\nonumber\\
\hspace{-7mm}
&+& \!\!\!\!\frac{(3s^3 - 3s^2 s' - 6s s'^2 - s'^3)}{12L} - \frac{(5s^2 -
12ss'- 4as'^2)}{24} - \frac{s' a}{6} L + \frac{(20a-3)}{720} L^2,\label{C6}\\
\hspace{-7mm}
J_3 (s, s')\!\!\! \!&=&\!\!\! \!\int^L_0 d t\,\Delta (t, s)\Delta (t, s') = -
\frac{(s^4 + 6s^2 s'^2 + s'^4)}{60L} + \frac{(s^2 + 3s'^2)s}{6}
\nonumber\\
\hspace{-7mm} & - &\!\!\!\! \frac{(s^2 + s'^2)}{6}L + \frac{(5a^2 -
10a + 6)}{45} L^3.\label{C7}\eeq

These are the building blocks for other relations
\beq \langle U_2 (s, s') \rangle &=& \frac{(d-1)}{2}\left[D_1 (s,
s') - \frac{(d+1)}{4} D_2^2 (s, s')\right] = - \frac{(d-1)(s -
s')}{4}\left[\frac{2La}{3}\right.\nonumber\\\hspace{-7mm} &+&
\left.\frac{(d-3)s - (d+1)s'}{2} - \frac{(d-1)s^2 - (d+1)s'^2}{L} +
\frac{d (s - s')(s + s')^2}{2L^2}\right]\label{C8},\\\hspace{-7mm}
K_1 (s, s') &=& J_1 (s, s') - \frac{1}{2}J_1 (s, s) - \frac{1}{2}
J_1 (s', s')\nonumber\\\hspace{-7mm} & = & - \,\,(s - s')
\left[\frac{La}{6} - \frac{(s + s')}{4} + \frac{(s^2 + s s' +
s'^2)}{6L}\right]\label{C9},\\\hspace{-7mm} K_2 (s, s') &=&J_2 (s,
s') + J_2 (s', s) - J_2 (s, s) - J_2 (s',
s')\nonumber\\\hspace{-7mm} & = & - \,\,(s - s')^2 \left[\frac{1}{4}
- \frac{(5s + 7s')}{12L} + \frac{(s +
s')^2}{4L^2}\right]\label{C10},\\ \hspace{-7mm} K_3 (s, s') &=& J_3
(s, s') - \frac{1}{2} J_3 (s, s) - \frac{1}{2} J_3 (s', s') = - (s -
s')^2 \left[\frac{(s + 2s')}{6} - \frac{(s +
s')^2}{8L}\right]\label{C11}\hspace{-2pt},\\ \hspace{-7mm} K_4 (s,
s') &=& \Delta (0, s)\Delta (0, s') - \frac{1}{2} \Delta^2 (0, s)  -
\frac{1}{2} \Delta^2 (0, s') = - \frac{(s - s')^2 (s + s' -
2L)^2}{8L^2}\label{C12},\\\hspace{-7mm} K_5 (s, s') &=& \Delta (L,
s)\Delta (L, s') - \frac{1}{2} \Delta^2 (L, s)  - \frac{1}{2}
\Delta^2 (L, s') = - \frac{(s^2 - s'^2)^2}{8L^2}.\label{C12bis}\eeq

\section{Relation with Distribution from Diffusion Equation}
\label{APPD}
The distribution function derived in
Ref.~\cite{largestiff} for $d = 3$ dimensions has the form
\beq  G (r; t) = {1\over 4\pi}\left[f_0 (t)\,\delta (r-1) + {f_1
(t)\over r}\,\delta' (r-1) + {f_2 (t)\over r}\,\delta'' (r-1) + {f_3
(t)\over r}\,\delta''' (r-1) + \dots \right]\,,\label{D1}\eeq
where the coefficients $f_i (t)$ are the polynomials in powers of
$t$.  To order $t^3$, their expansions read
\beq f_0 (t)& = & 1 + {1\over 3}\,t + {1\over 15}\,t^2 + {4\over
315}\,t^3 + \dots~,\label{D2}\\
f_1 (t)& = & f_0 (t) - 1\,,\label{D3}\\
f_2 (t)& = & {7\over 90}\,t^2 - {1\over 630}\,t^3 + \dots~,
\label{D4}\\
f_3 (t)& = & {31\over 1890}\,t^3 - \dots~.\label{D5}\eeq
Depending on the reduced distance $r$, the expression~(\ref{D1})
cannot be directly comparable with the distribution
function~(\ref{6.16}). To remove this, we expand
\beq {1\over r} = {1\over (r-1) + 1} = 1 - (r-1) + (r-1)^2 - (r-1)^3
+ \dots\,,\label{D6}\eeq
Substituting the expansion~(\ref{D6}) into Eq.~(\ref{D1}), we use
the properties of $\delta$-function
\beq
x\,\delta' (x) &=& - \delta (x)\,,\, x^n \,\delta' (x) \equiv 0\,, n > 1\,,\label{D7}\\
x\,\delta'' (x) &=& - 2\delta' (x)\,,\, x^2\,\delta'' (x) = 2\delta
(x)\,,\, x^n \,\delta' (x) \equiv
0\,, n > 2\,,\label{D8}\\
x\,\delta''' (x) &=& - 3\delta'' (x)\,,\, x^2\,\delta''' (x) =
6\delta'(x)\,,\, x^3\,\delta''' (x) = - 6\delta(x)\,, \, x^n
\,\delta' (x) \equiv 0\,, n > 3\,.\label{D9}\eeq
This brings the distribution function~(\ref{D1}) in the form
(\ref{6.16}) with
\beq
p_0 (t) & = & 2f_0 (t) + 2\,{f_2 (t)\over t} + 6\,{f_3 (t)\over
t^2} - 1 + \dots = 1 + {2\over 3}\,t + {13\over 45}\,t^2 + {38\over
315}\,t^3 + \dots~,\label{D10}\\
p_1 (t) & = & f_0 (t) + 2\,{f_2 (t)\over t} + 6\,{f_3 (t)\over t^2}
- 1 + \dots = {1\over 3}\,t + {2\over 9}\,t^2 + {34\over 315}\,t^3 +
\dots~,\label{D11}\\
p_2 (t) & = & \,{f_2 (t)\over t} + 3\,{f_3 (t)\over t^2} + \dots =
{7\over 90}\,t^2 + {1\over 21}\,t^3 + \dots~,\label{D12}\\
p_3 (t) & = & {f_3 (t)\over t^2} + \dots = {31\over 1890}\,t^3 +
\dots~.\label{D13}\eeq
%


\end{document}